\documentclass[intlimits,twoside,a4paper]{article}
\usepackage{amsmath,amssymb}
\usepackage{graphicx}
\usepackage{wrapfig}
\usepackage[T2A]{fontenc}
\usepackage[cp1251]{inputenc}
\usepackage{array}
\usepackage{color}

\usepackage[eqsecnum]{cmpj}

\issue{2011}{14}{3}{33801}

\doinumber{10.5488/CMP.14.33801}

\title[Microphase formation under the effect of a matrix]%
{Two-dimensional systems with competing interactions: microphase formation under the
effect of a disordered porous matrix}
\author[D.F. Schwanzer, G. Kahl]{D.F.~Schwanzer, G.~Kahl}

\address{Institut f\"ur Theoretische Physik and Center
  for Computational Materials Science (CMS), Technische Universit\"at
  Wien, Wiedner Hauptstra{\ss}e 8--10, A--1040 Wien, Austria}
\date{Received May 18, 2011, in final form July 4, 2011}

\authorcopyright{D.F. Schwanzer, G. Kahl, 2011}

\begin{document}

\maketitle

\begin{abstract}
We have investigated the effect of a disordered porous matrix on
the cluster microphase formation of a two dimensional system where
fluid particles interact via competing interactions. To this end we have
performed extensive Monte Carlo simulations and have systematically
varied the densities of the fluid and of the matrix as well as the
interaction between the matrix particles and between the matrix and
fluid particles. Our results provide evidence that the matrix \textit{does} 
have a distinct effect on the microphase formation of the
fluid particles: as long as the matrix particles interact both among
themselves as well as with the fluid particles via a simple hard
sphere potential, they essentially reduce the available space, in
which the fluid particles form a cluster microphase. On the other
hand, if we turn on a long-range tail in the matrix-matrix and in
the matrix-fluid interactions, the matrix particles become nucleation
centers for the clusters formed by the fluid particles.
\pacs 82.70.Dd, 46.65.+g, 47.54.Bd, 61.20.Ja
\end{abstract}

\section{Introduction}
\label{sec:introduction}

In recent years, systems with so-called competing interactions have
attracted considerable interest in the scientific community. In such
systems, the (effective) potential $\Phi(r)$ between two particles
that are separated by a distance $r$ is given by a strongly repulsive
core region and an adjacent tail which is characterized by
interparticle interactions operating on different length scales: it has an attractive component at short distances
and a repulsive component at intermediate distances, eventually
tending to zero for $r \to \infty$. In the past, experimental
investigations~\cite{Klo06, Gel99FAR, Eli97, Ghe97, Seu95} on such
systems have been complemented by thorough theoretical studies~\cite{Imp04, Imp06, Pin06, Imp07, Imp08, Arc08PRE, Imp06WS, Arc07PRE,
  Arc07JCP, Arc08JCPM, Lee10, Sch10}, most of which have focused on
a two dimensional case.

The most remarkable feature of such systems is their capability of
self-organizing at sufficiently low temperatures -- despite a
spherically symmetric potential~-- into so-called microphases, i.e.,
highly inhomogeneous, ordered patterns. In two dimensions, on which
the present contribution will focus,  as the density increases, the following morphologies are
reported: clusters, stripes and bubbles (or
inverse clusters). The aforementioned theoretical investigations
devoted to the two dimensional case~\cite{Imp04, Imp06, Pin06,
  Imp07, Imp08, Arc08PRE, Imp06WS, Arc07PRE, Sch10} have also unveiled
other interesting and surprising features of these systems, related to
the phase behaviour and to the properties of their disordered
phases. However, since these aspects are of less relevance to the present contribution
 we refer the reader to these
references for more details.

In the present contribution we investigate the following scenario: we
consider a disordered matrix, generated by a frozen (quenched)
equilibrium configuration of matrix particles and immerse fluid particles (annealed component) into this
configurations that interact via
potentials operating on different length scales. Using Monte Carlo
(MC) simulations we study the effect of the matrix on the
microphase formation scenario of the fluid particles, varying the
fluid and the matrix densities as well as the fluid-matrix and the
matrix-matrix interactions. In these investigations we focus on a
region in phase space, where the equilibrated fluid forms clusters,
i.e., we concentrate on sufficiently low temperatures and on rather low fluid
densities, $\rho_{\rm f}$. In order to generate such systems we have
used the quenched-annealed concept (see, e.g.,~\cite{Kur09, Kur10} and
references therein): in a first step, we have generated an equilibrium configuration
of the matrix particles in a MC
simulations at a desired matrix density $\rho_{\rm m}$. Keeping the positions of the matrix particles
fixed we have then immersed a fluid of a density $\rho_{\rm f}$ into
the system. Using extensive MC simulations we have evaluated, via
suitable averaging processes (see, e.g.,~\cite{Kur09, Kur10}),
the information on the static structure of the fluid and on the
mobility of the fluid particles. This information provides us, in
combination with a visual inspection of the snapshots, with information
on the microphase formation of the fluid particles under the
external field of the matrix particles.

We summarize our observations as follows: both the densities
$\rho_{\rm f}$ and $\rho_{\rm m}$ as well as the type of interaction
between fluid-matrix and matrix-matrix particles have a distinct
effect on the observed microphase formation scenarios of the fluid
particles: at small matrix densities, the cluster formation is
essentially unaffected by the presence of the matrix
particles. However, as $\rho_{\rm m}$ increases, the interparticle
potentials both between the fluid and the matrix as well as between
the matrix particles becomes relevant: (i) hard sphere matrix
particles can foster or suppress microphase formation, depending on
whether they leave sufficient space for the fluid particles to
self-assemble in clusters or not; (ii) on the other hand, if the
fluid-matrix interaction is essentially of the same type as the
fluid-fluid interaction, then we observe that the matrix particles act
as nucleation sites for the clusters formed by the fluid particles.

The paper is organized as follows: in the subsequent section we
present our model system and summarize the details concerning the MC
simulations. In section 3 we discuss our results and close the paper
with concluding remarks.

\section{Model and theoretical approach}
\label{sec:model_theory}

As in our previous investigation~\cite{Sch10}, the fluid particles
(index `f') interact via a spherically symmetric potential $\Phi_{\rm
  ff}(r) = \Phi_{\rm IR}(r)$, originally proposed in this
parametrization by Imperio and Reatto (IR)~\cite{Imp04}; it is given
by
\begin{equation}
\label{eq:potential}
\Phi_{\rm IR}(r) = \left \{ \begin{array} {l@{~~~~~~}l}
                   \infty, & r \leqslant \sigma, \\
                   -\epsilon_{\rm a} \frac{\sigma^{2}}{R_{\rm a}^{2}} \exp \left( -\frac{r}{R_{\rm a}} \right)
                   +\epsilon_{\rm r} \frac{\sigma^{2}}{R_{\rm r}^{2}} \exp \left( -\frac{r}{R_{\rm r}} \right),
                        & r > \sigma. \\
                   \end{array}
                   \right.
\end{equation}
$\sigma$ is the diameter of the impenetrable hard-core region,
$\epsilon_{\rm a}$ ($\epsilon_{\rm r}$) and $R_{\rm a}$ ($R_{\rm r}$)
represent the strength and the range of the attractive (repulsive)
contributions to the potential tail of $\Phi_{\rm IR}(r)$ for $r \geqslant \sigma$, respectively.  Throughout we have used $\epsilon_{\rm a} =
      {\epsilon_{\rm r}}$ and $R_{\rm a} = \sigma$ and
      $R_{\rm r} = 2 \sigma$. Further we introduce the temperature $T$
      and the fluid and matrix (area) densities $\rho_{\rm f}$ and
      $\rho_{\rm m}$, respectively. Further we will use reduced
      units, i.e., $r^\ast = r/\sigma$, $k^\ast = k \sigma$, and
      $\rho^\ast = \rho \sigma^2$; for the temperature $T$ we have
      used the arbitrary temperature scale introduced in~\cite{Imp04},
      namely that for $R_{\rm r}/R_{\rm a} = 2$ and $\epsilon_{\rm
        r}/\epsilon_{\rm a} = 1$, $\Phi(\sigma)/k_{\rm B} T = -1$. For
      simplicity we will drop henceforward the asterisk.

For the interactions between the matrix particles (index `m'),
$\Phi_{\rm mm}(r)$, and the cross interaction between fluid and matrix
particles, $\Phi_{\rm fm}(r)$, we have used either a simple hard (HS)
sphere potential, $\Phi_{\rm HS}(r)$ (with a HS diameter $\sigma$) or
the IR potential, specified in equation (\ref{eq:potential}).  In this
way we are able to study the effect of the matrix on the microphase
separation scenario of the fluid particles, including thereby both
simple excluded volume effects [i.e., when $\Phi_{\rm fm}(r) =
  \Phi_{\rm HS}(r)$] and energetic effects [i.e., when
  $\Phi_{\rm fm}(r) = \Phi_{\rm IR}(r)$]. To be more specific, the
fluid particles {\it always} interact via $\Phi_{\rm IR}(r)$; for the
fluid-matrix and the matrix-matrix interactions we have considered the
following three combinations (cases 1 to 3):
\begin{itemize}
\item {\bf case 1}: $\Phi_{\rm fm}(r) = \Phi_{\rm HS}(r)$ and
  $\Phi_{\rm mm}(r) = \Phi_{\rm HS}(r)$;
\item {\bf case 2}: $\Phi_{\rm fm}(r) = \Phi_{\rm IR}(r)$ and
  $\Phi_{\rm mm}(r) = \Phi_{\rm HS}(r)$;
\item {\bf case 3}: $\Phi_{\rm fm}(r) = \Phi_{\rm IR}(r)$ and
  $\Phi_{\rm mm}(r) = \Phi_{\rm IR}(r)$.
\end{itemize}

\begin{figure}[ht]
\begin{center}
\includegraphics[width=0.50\textwidth]
          {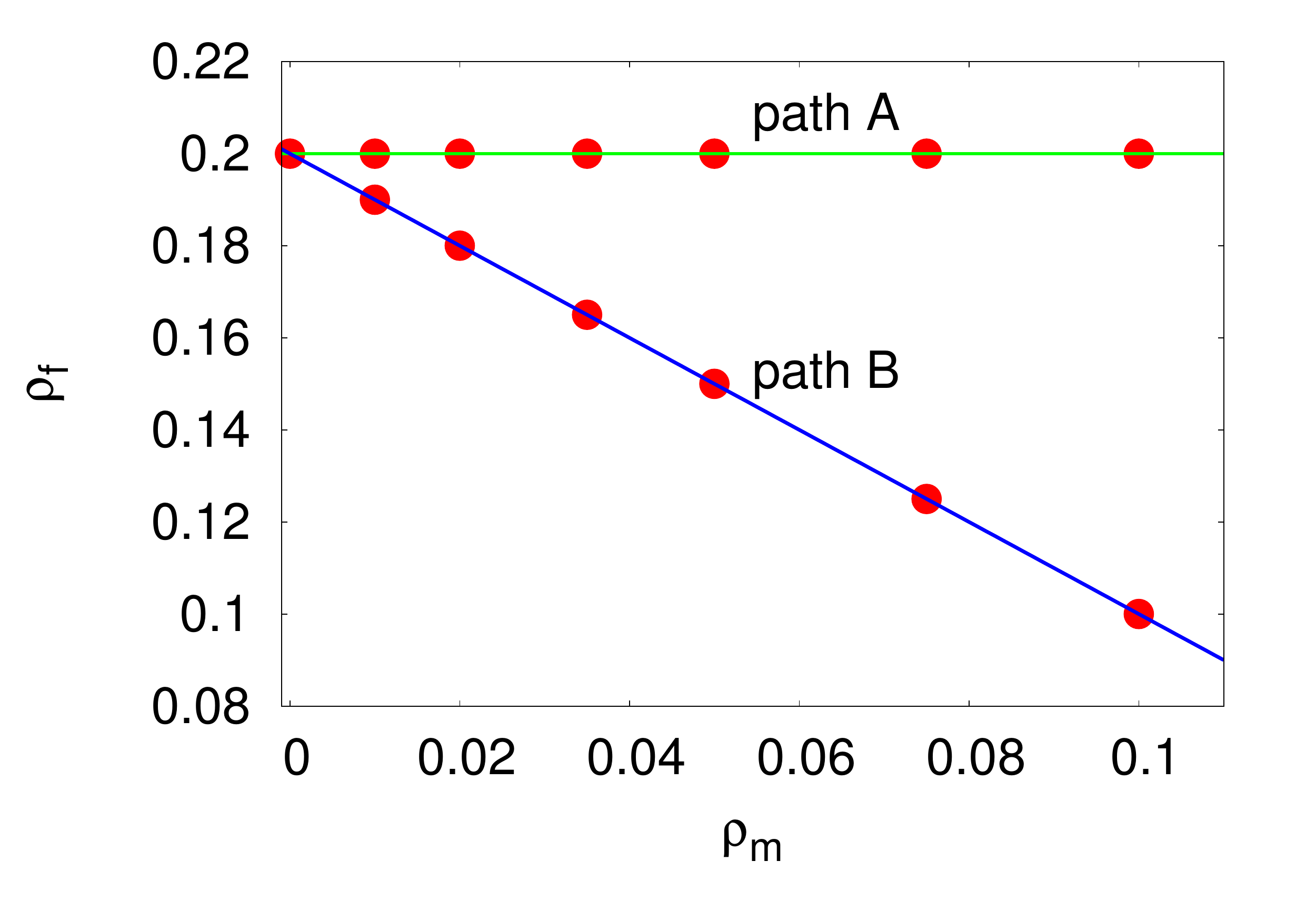}
\end{center}
\caption{(color online) Symbols in the $(\rho_{\rm f}, \rho_{\rm
    m})$-plane represent state points that have been investigated in
  this contribution. Along path $A$, the fluid density is fixed to
  $\rho_{\rm f} = 0.2$, while the matrix density $\rho_{\rm m}$ varies
  from 0 to 0.1. Along path $B$, the total density, $\rho_{\rm t} =
  \rho_{\rm f}+ \rho_{\rm m}$, is kept fixed to $\rho_{\rm t} = 0.2$:
  $\rho_{\rm m}$ increases from 0 to 0.1, i.e. $\rho_{\rm f}$
  decreases from 0.2 to 0.1.}
\label{fig:pathways}
\end{figure}

The range of the densities $\rho_{\rm f}$ and $\rho_{\rm m}$ has been
restricted to smaller values below 0.2 and, throughout, temperature
was set to $T = 0.5$, corresponding to the region in phase space where
cluster formation is expected to occur for the pure fluid (cf. figure~4 in~\cite{Sch10}). In figure~\ref{fig:pathways} we have specified in
the $(\rho_{\rm f}, \rho_{\rm m})$-plane those systems that we have
investigated in this contribution: for state points located along the path
$A$, $\rho_{\rm f}$ is kept fixed to a value of 0.2, while $\rho_{\rm
  m}$ varies from 0 up to 0.1; along the path $B$, the total density,
$\rho_{\rm t} = \rho_{\rm f} + \rho_{\rm m}$, is kept fixed to a value
of 0.2.

System properties have been investigated via standard NVT MC
simulations. $N_{\rm f}$ fluid and $N_{\rm m}$ matrix particles have
been considered in a square box, using periodic boundary
conditions. Depending on the total density, from 2~000 to 6~000
particles have been considered. Computational speed-up was achieved by
refined techniques described more in detail in subsection 2.3 of~\cite{Sch10}. {The potential~(\ref{eq:potential}) was
  truncated at $r_{\rm cut} = 17.2$, for the maximum displacement we
  used throughout a value of 0.6.}  Matrix configurations were created
in simulation runs of an equilibrated system of matrix particles at a
density $\rho_{\rm m}$. At several instances, the simulation has been
halted and the positions of the particles were recorded; thereby
different but equivalent matrix configurations were
produced. Simulations have been extended over 1~000~000 MC sweeps. For
a given state point, characterized by $\rho_{\rm f}$ and $\rho_{\rm
  m}$, observables were obtained in a two-step averaging procedure:
in a first step, an average was taken over the degrees of freedom of
the fluid particles for a given matrix configuration (involving 10~000
independent particle configurations); in a second step, these results
were averaged over five independent matrix configurations.

\section{Results}
\label{sec:results}

We discuss the results obtained for our system along the following
lines: after a {\it qualitative} visual inspection of the selected
representative snapshots obtained from MC simulations, we will analyse
these findings on a more {\it quantitative} level by studying the pair
structure and the mean square displacement.

\subsection{Snapshots}
\label{subsec:snapshots}

\begin{figure}[htpb]
\begin{tabular}{m{0.1\textwidth}m{0.35\textwidth}m{0.35\textwidth}}
& \multicolumn{2}{c}{\includegraphics[width=0.3\textwidth]
          {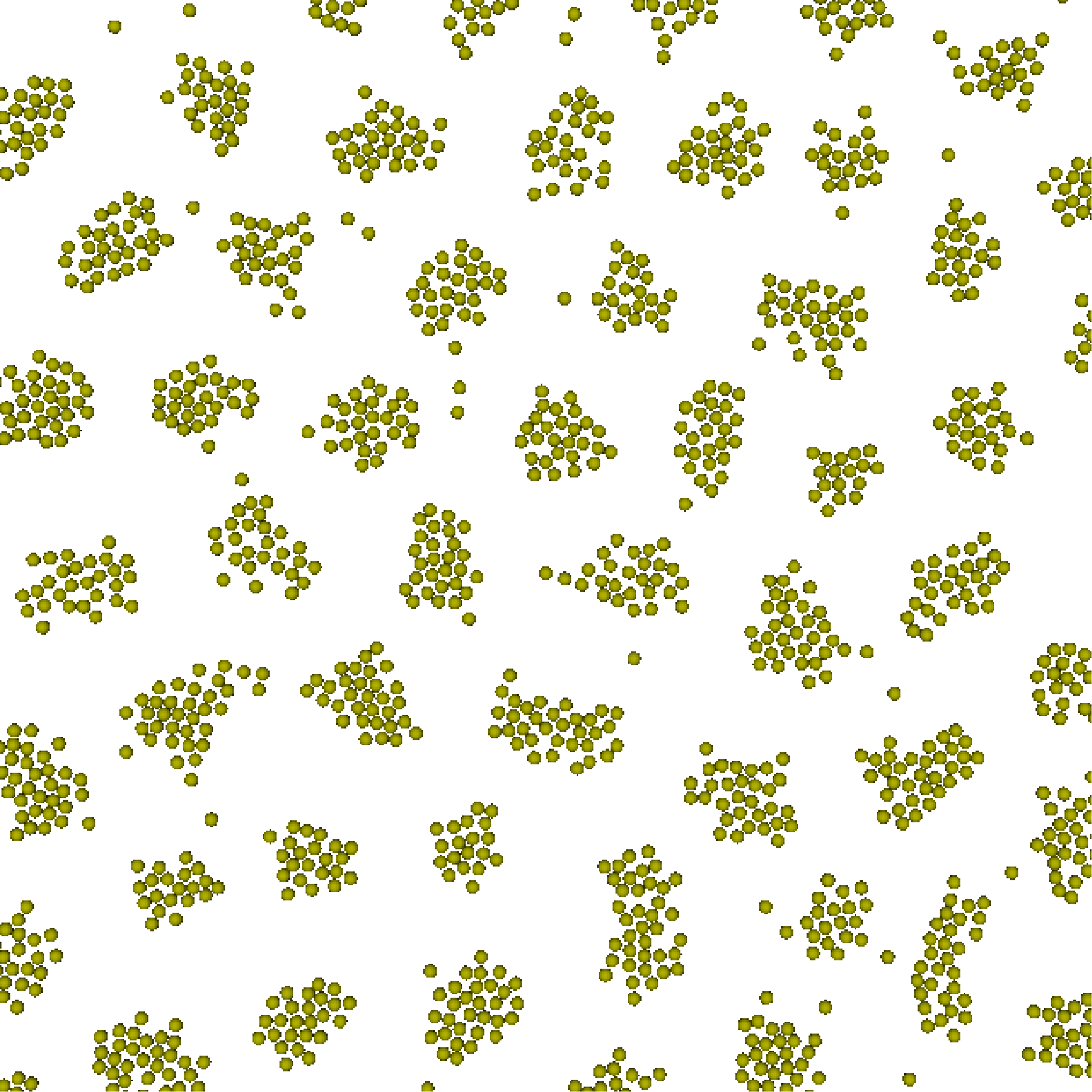}} \\
{\bf case 1} &
\includegraphics[width=0.3\textwidth]
          {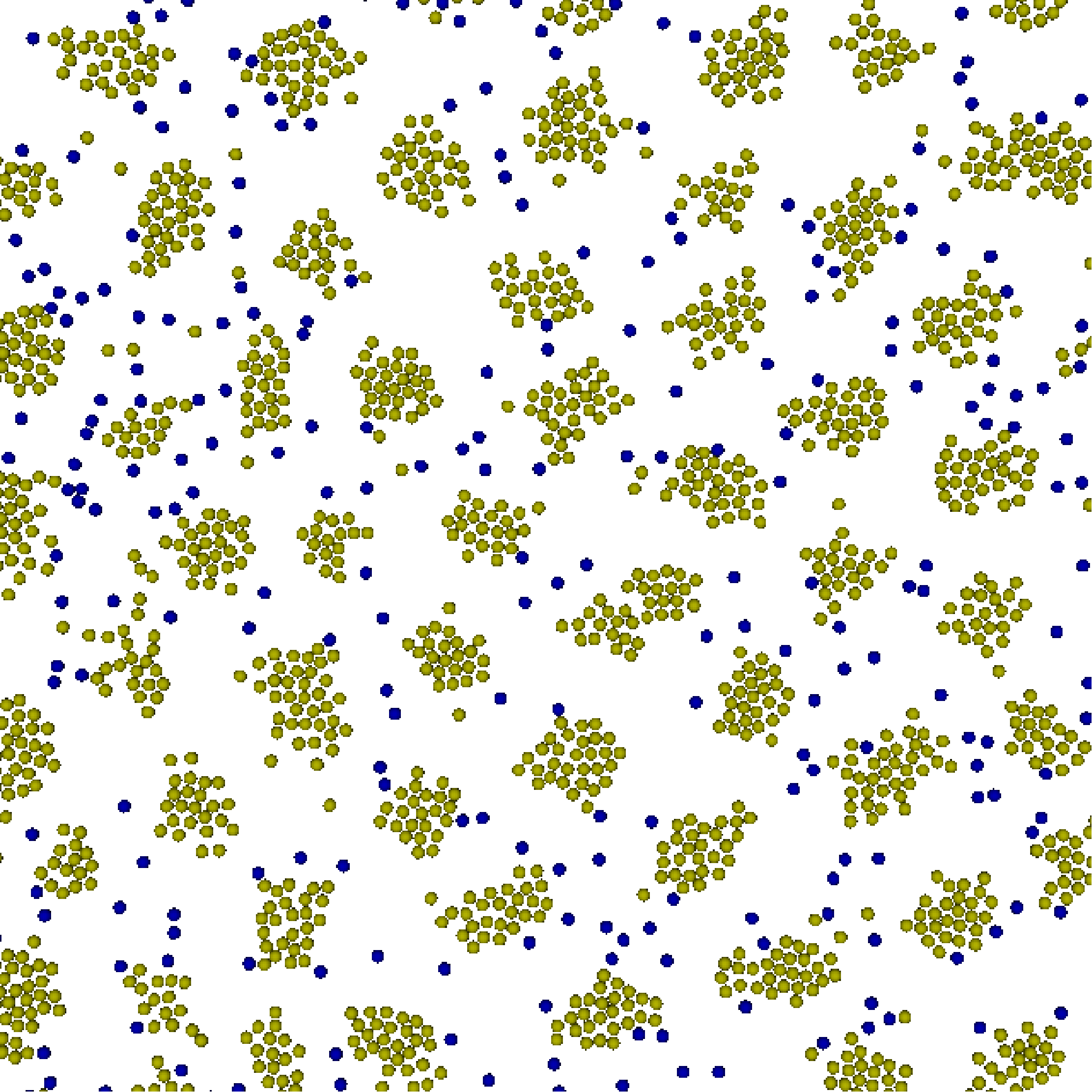} &
\includegraphics[width=0.3\textwidth]
          {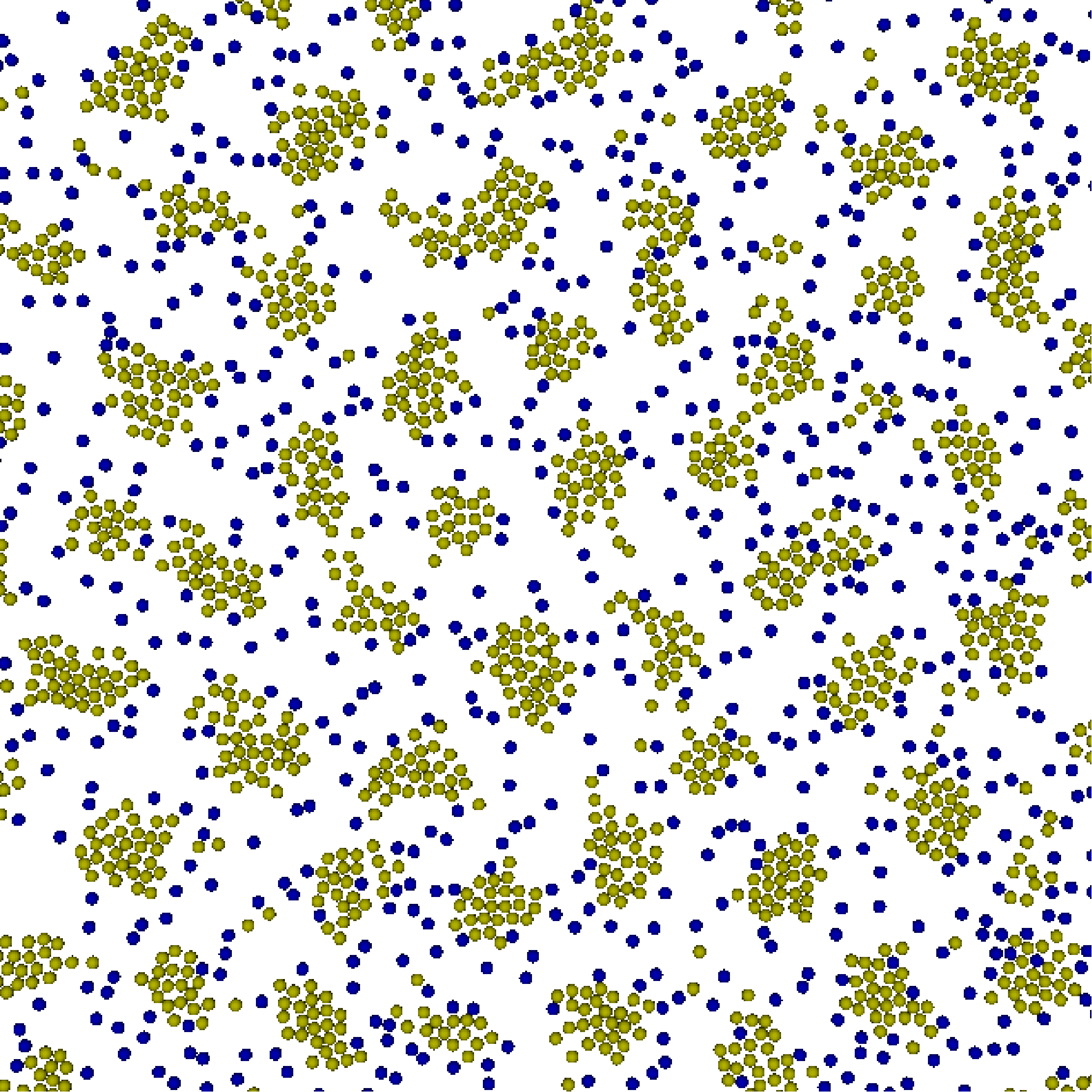} \\
{\bf case 2} &
\includegraphics[width=0.3\textwidth]
          {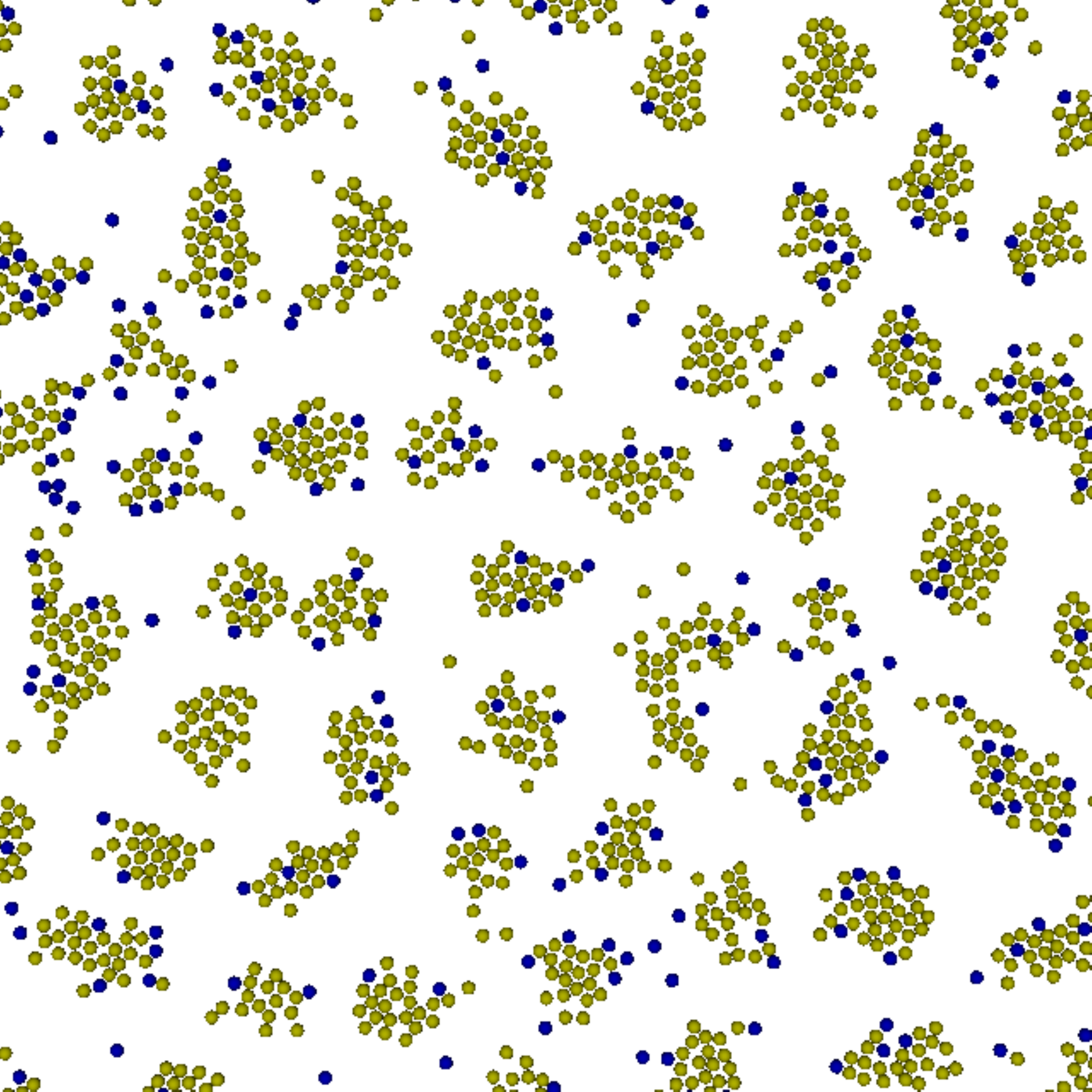} &
\includegraphics[width=0.3\textwidth]
          {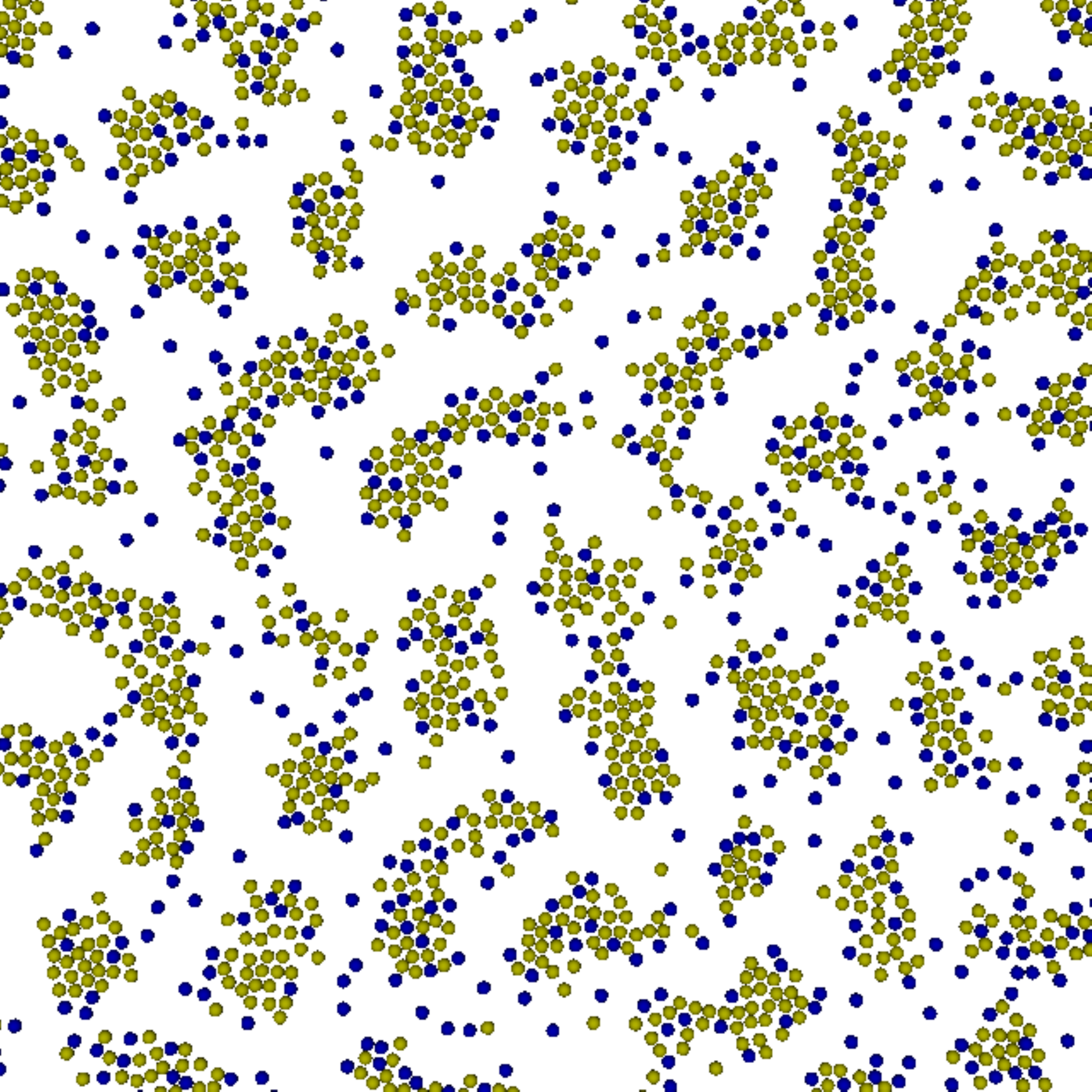} \\
{\bf case 3} &
\includegraphics[width=0.3\textwidth]
          {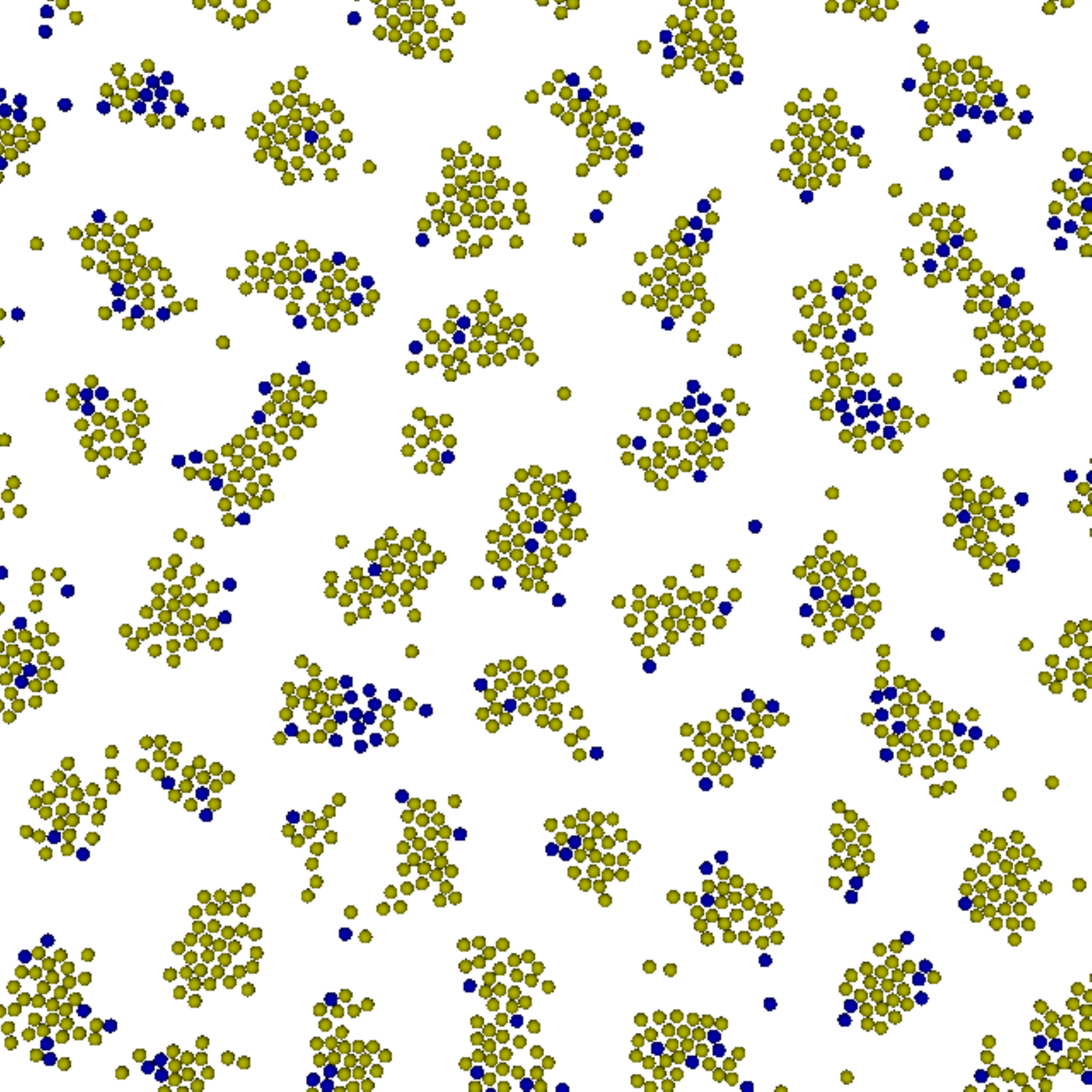} &
\includegraphics[width=0.3\textwidth]
          {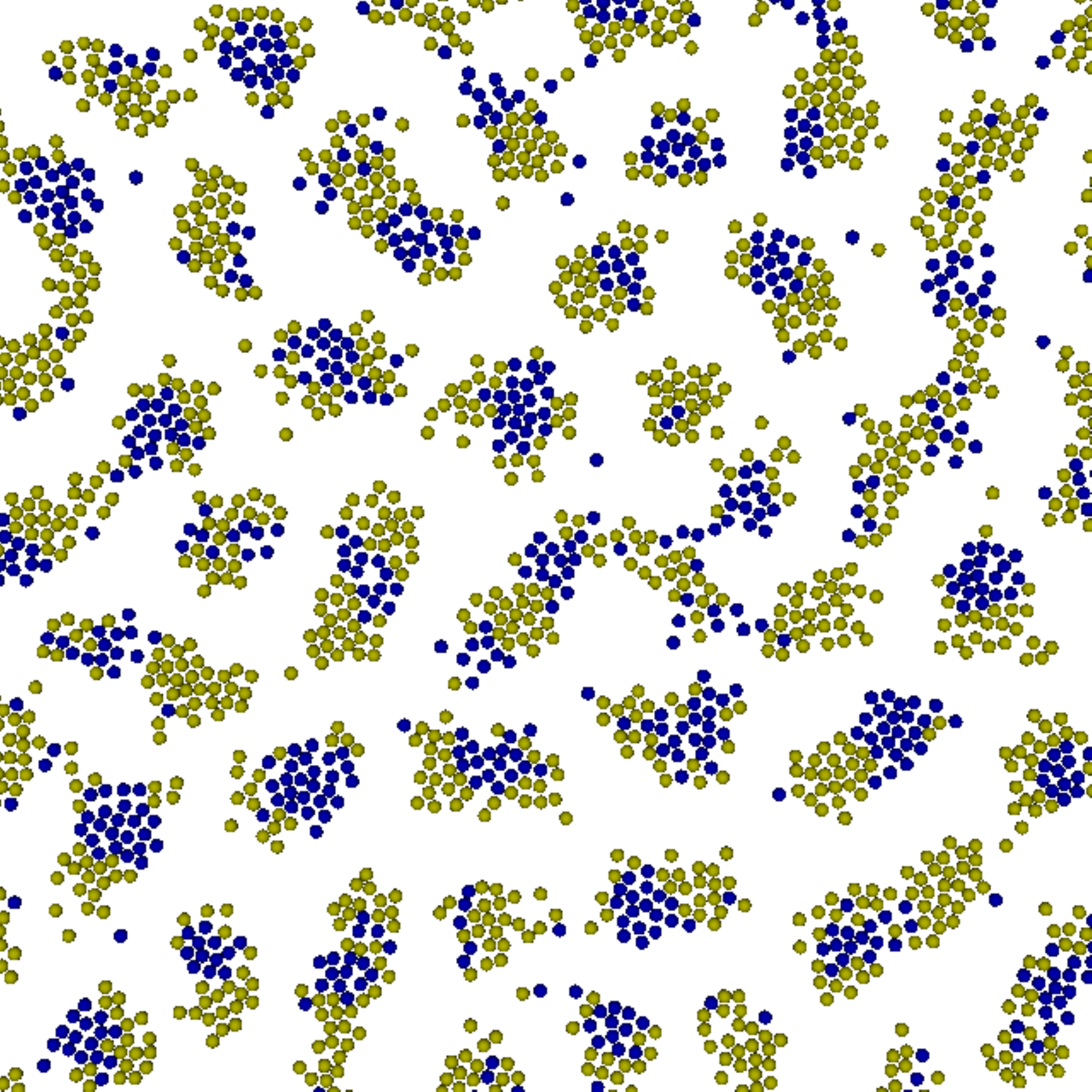} \\
\end{tabular}
\caption{(color online) Top panel: snapshot of an equilibrated fluid
  at $\rho_{\rm f} = 0.2$. Other panels: selected, representative
  snapshots of systems located in the $(\rho_{\rm f}, \rho_{\rm
    m})$-plane along path $A$, spe\-ci\-fied in figure~\ref{fig:pathways}. The three rows correspond to cases 1 to 3, as
  specified in the text, the two columns correspond to a matrix
  density $\rho_{\rm m} = 0.035$ (left column) and $\rho_{\rm m} =
  0.1$ (right column), respectively. Light (yellow) particles -- fluid
  particles, dark (blue) particles -- matrix particles.}
\label{fig:snapshots_A}
\end{figure}

In figure~\ref{fig:snapshots_A} we show selected representative
snapshots for systems located in the $(\rho_{\rm f}, \rho_{\rm
  m})$-plane along path $A$, specified in figure~\ref{fig:pathways},
i.e. starting with the pure fluid of density $\rho_{\rm f} = 0.2$
(i.e., $\rho_{\rm m} = 0$) and -- while keeping $\rho_{\rm f}$ fixed~-- increasing continuously the matrix density $\rho_{\rm m}$ from 0 to
0.1. We recall that -- according to the phase diagram depicted in
figure~4 of~\cite{Sch10} -- for this density-range and for the assumed
temperature, the matrix particles do not form clusters.
\begin{figure}[htpb]
\begin{tabular}{m{0.1\textwidth}m{0.35\textwidth}m{0.35\textwidth}} \\
& \multicolumn{2}{c}{\includegraphics[width=0.3\textwidth]
          {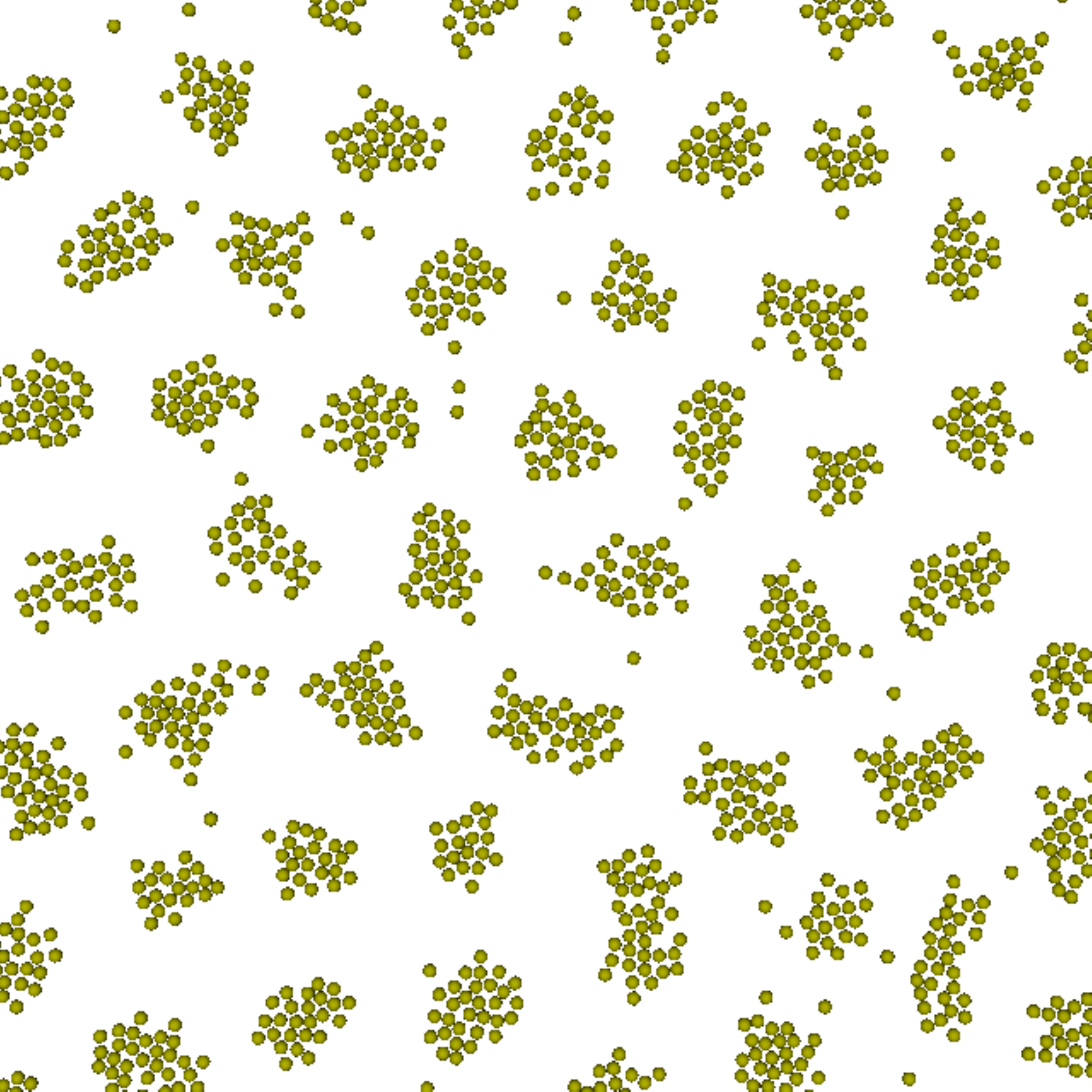}} \\
{\bf case 1} &
\includegraphics[width=0.3\textwidth]
          {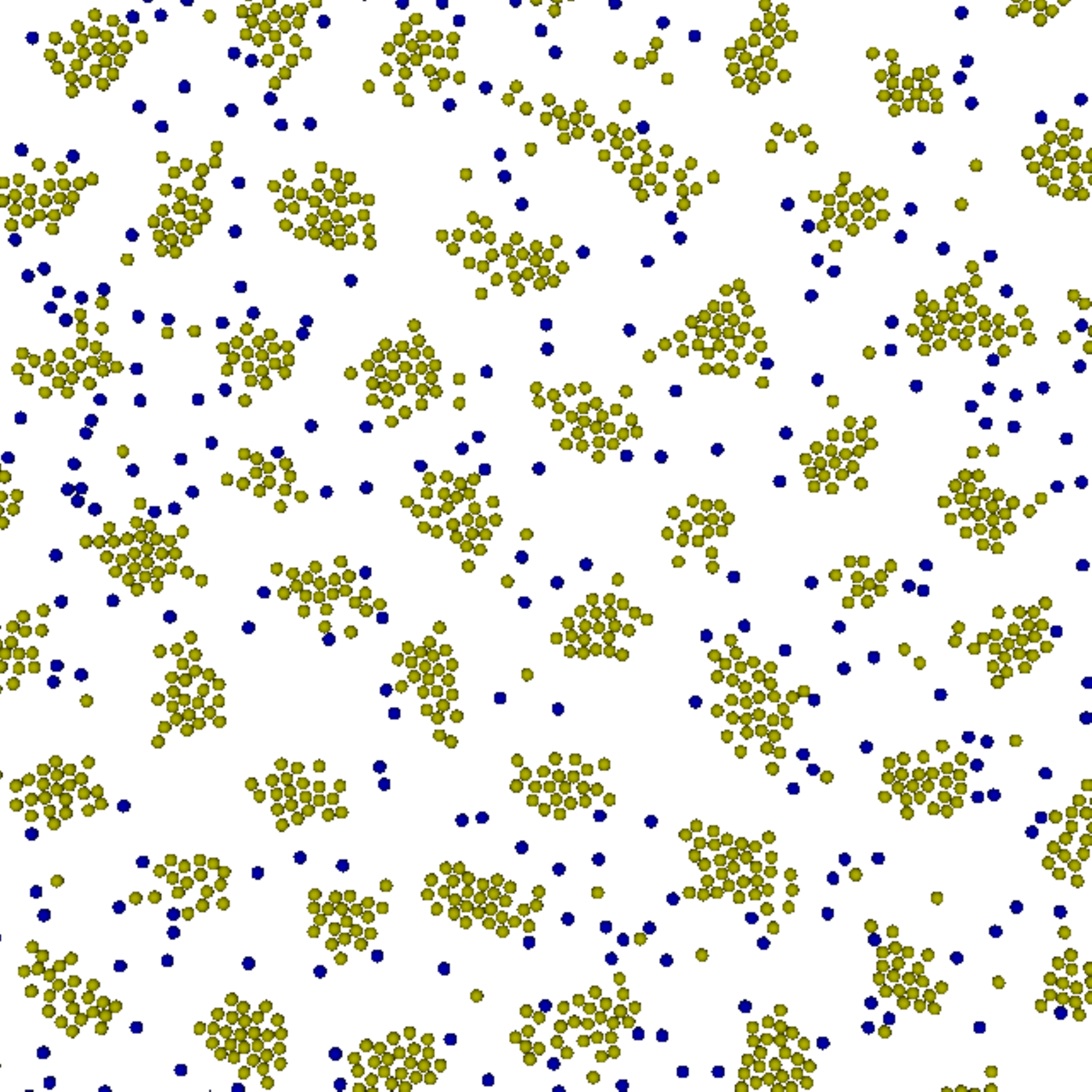} &
\includegraphics[width=0.3\textwidth]
          {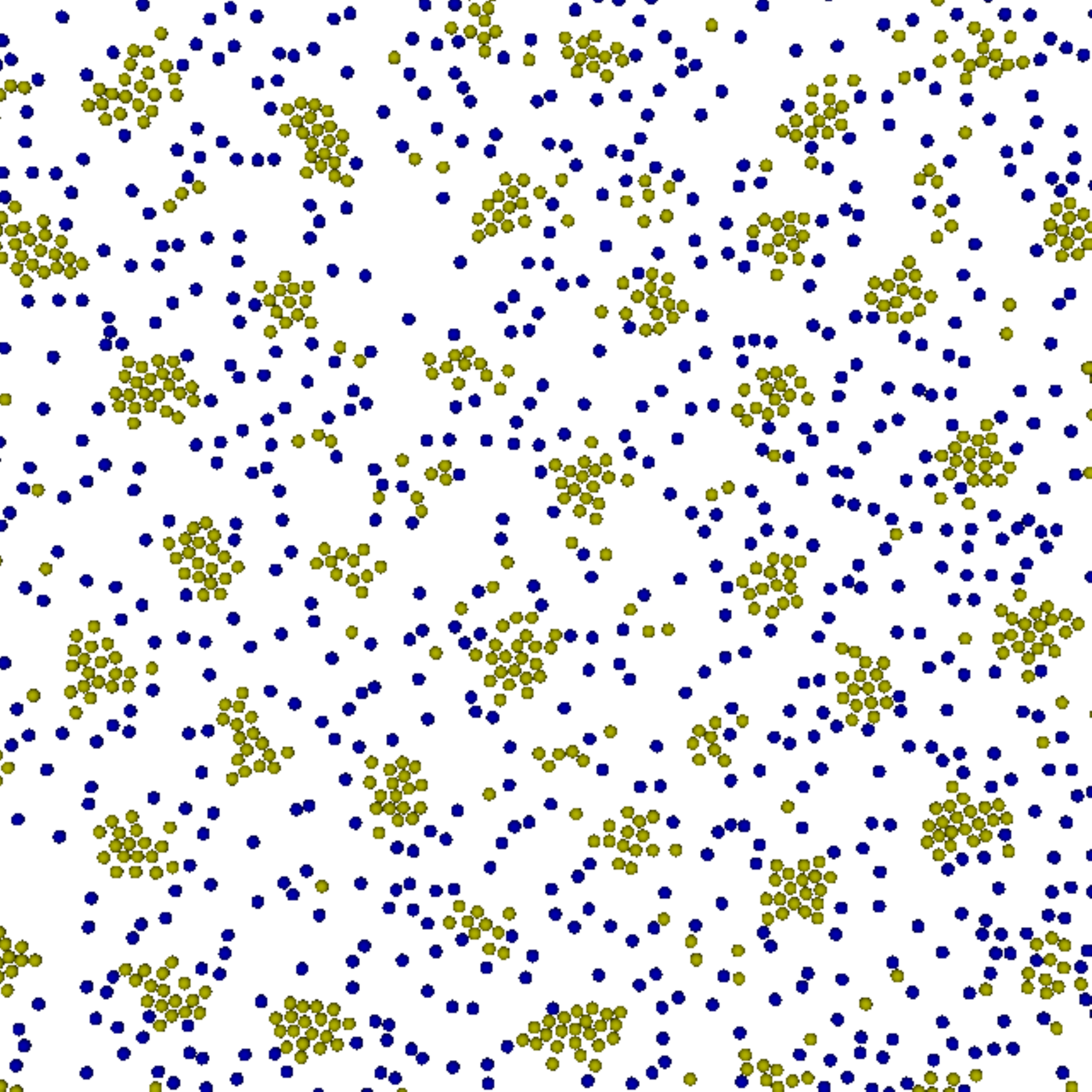} \\
{\bf case 2} &
\includegraphics[width=0.3\textwidth]
          {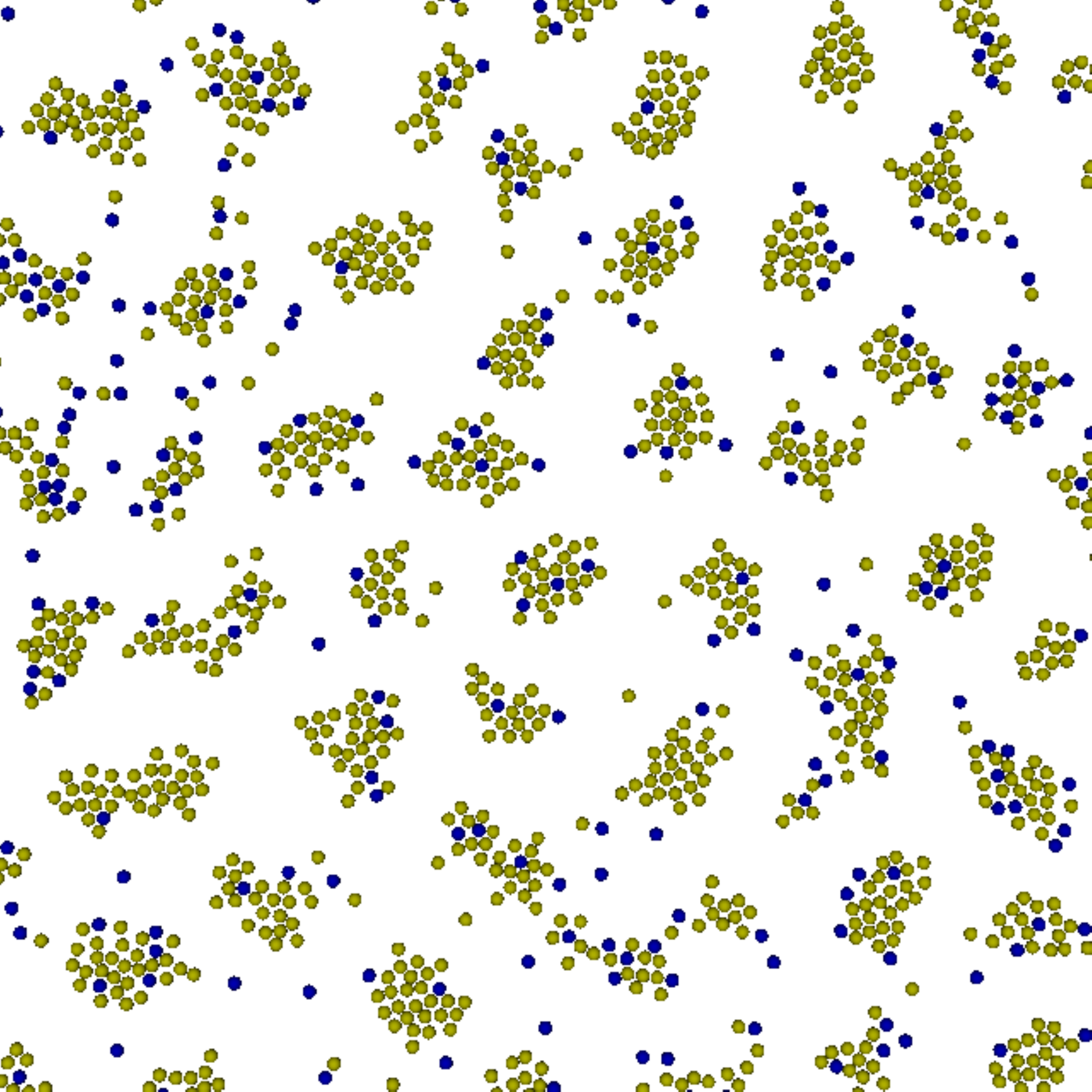} &
\includegraphics[width=0.3\textwidth]
          {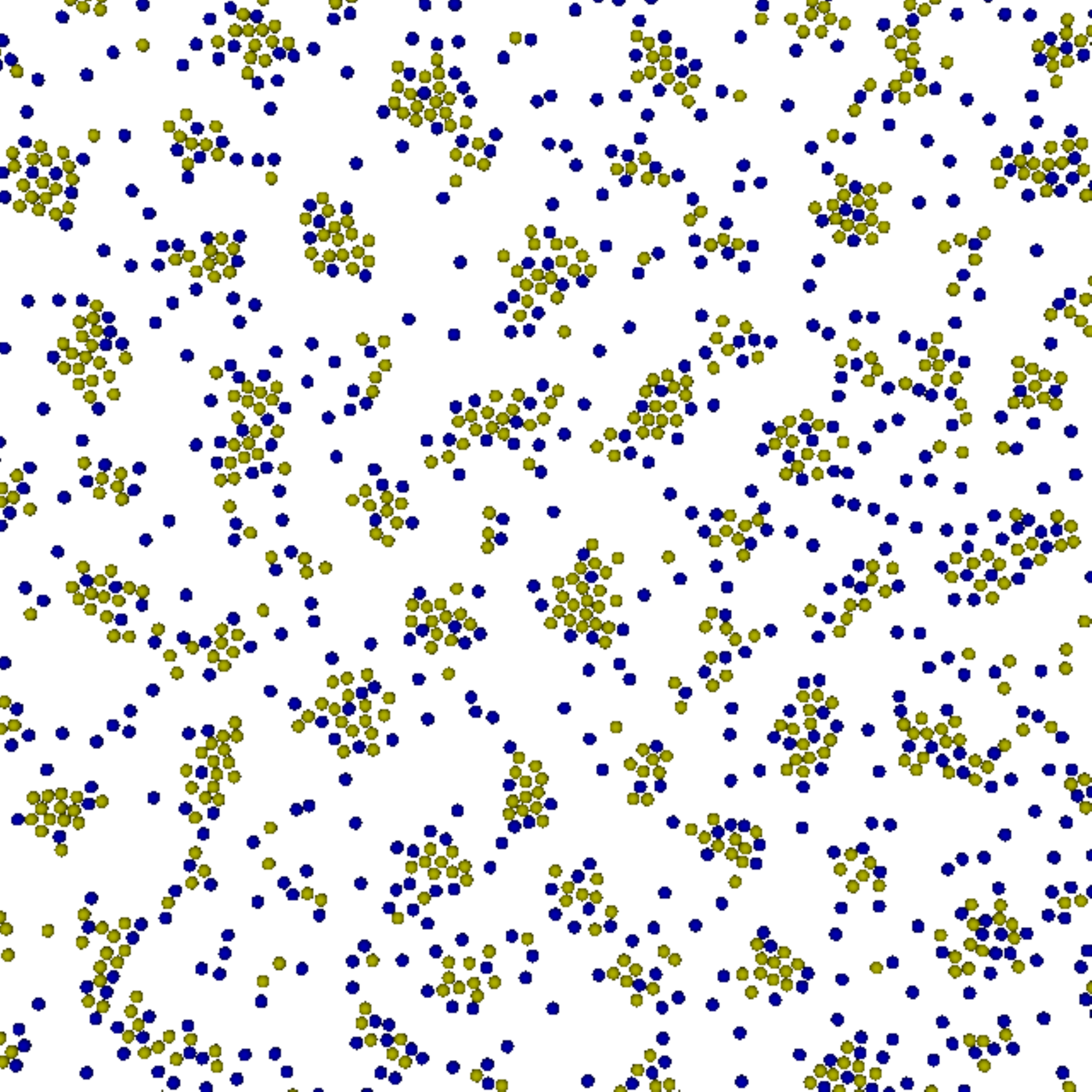} \\
{\bf case 3} &
\includegraphics[width=0.3\textwidth]
          {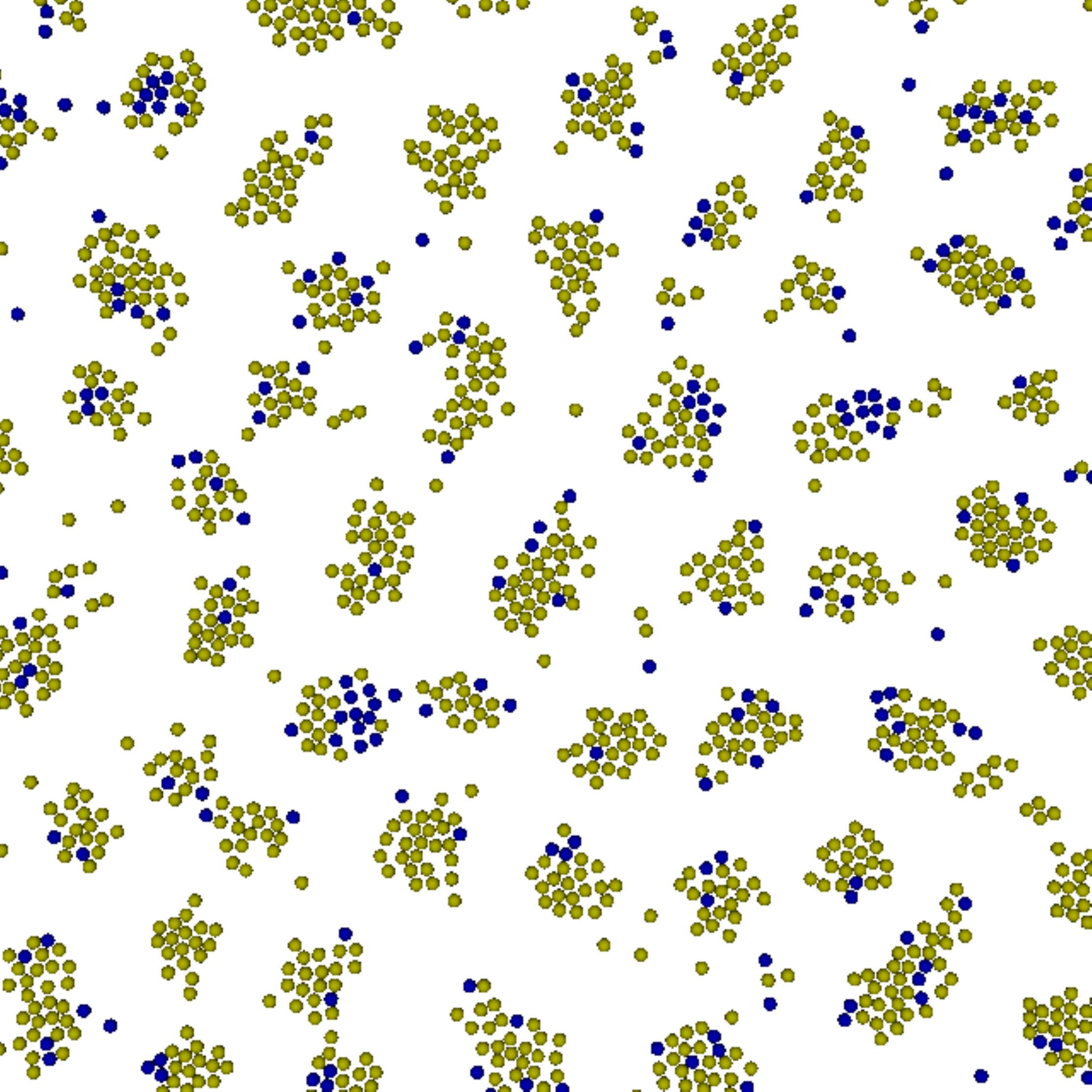} &
\includegraphics[width=0.3\textwidth]
          {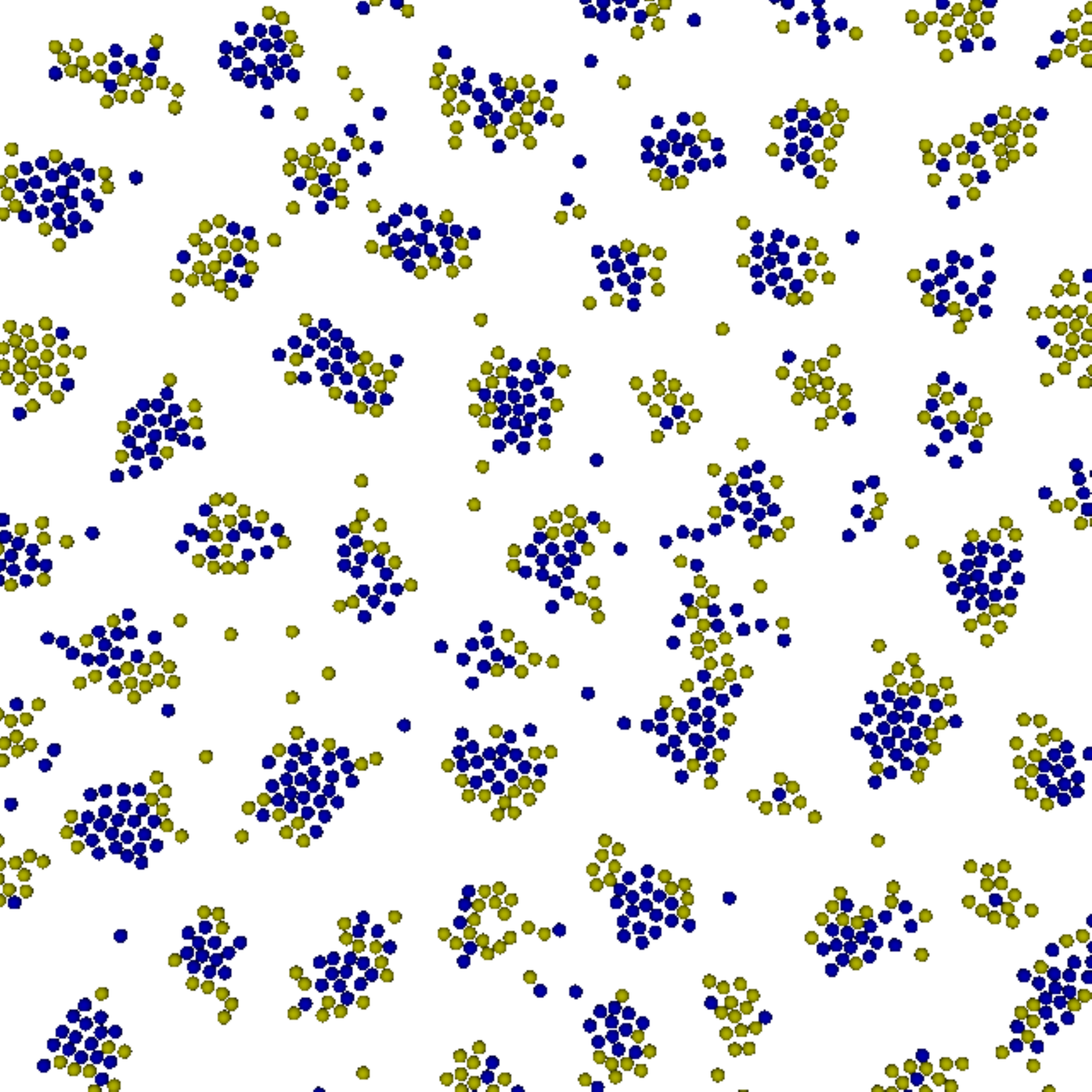} \\
\end{tabular}
\caption{(color online) Top panel: snapshot of an equilibrated fluid
  at $\rho_{\rm f} = 0.2$. Other panels: selected, representative
  snapshots of systems located in the $(\rho_{\rm f}, \rho_{\rm
    m})$-plane along path $B$, specified in figure~\ref{fig:pathways}. The three rows correspond to cases 1 to 3, as
  specified in the text, the two columns correspond to a matrix
  density $\rho_{\rm m} = 0.035$, i.e., $\rho_{\rm f} = 0.165$ (left
  column) and $\rho_{\rm m} = 0.1$, i.e., $\rho_{\rm f} = 0.1$ (right
  column), respectively. Light (yellow) particles -- fluid particles,
  dark (blue) particles -- matrix particles.}
\label{fig:snapshots_B}
\end{figure}
In the top
panel of figure~\ref{fig:snapshots_A} a typical equilibrium
configuration of the fluid (i.e., $\rho_{\rm m} = 0$) is depicted. We
start the discussion of the snapshots with three panels of the
left column where we consider a matrix at density $\rho_{\rm m} =
0.035$. As we now immerse the fluid into the quenched matrix
configuration we observe only a weak effect of the immobile matrix
particles on the formation of clusters by the fluid particles: from a
qualitative visual inspection of the snapshots we learn that the
number of clusters as well as their size is essentially the same for
three different interaction scenarios considered in cases 1 to
3. We only notice that for case 1, the essentially inert matrix
particles are located predominantly outside the clusters, while in
case 3, i.e. when both types of particles interact via IR potentials,
the matrix particles have become part of the clusters. This means that
in this case the matrix particles act as nucleation centers for the
clusters formed by the fluid particles.  We now repeat this experiment
at a higher matrix density, i.e. at $\rho_{\rm m} = 0.1$. Now the
matrix particles {\it do} have a distinct effect on the cluster
formation of the fluid particles. In case 1 the rather compact (i.e.,
close-to-spheric) clusters of the fluid particles are formed nearly
exclusively in those regions which have been left void by the matrix
particles. As we proceed to case 2, i.e. as we switch on an IR-tail in
the matrix-fluid particle interaction, the situation changes
drastically: due to a rather high matrix density, the clusters are
forced to arrange predominantly `around' the matrix particles and
with respect to case 1, the clusters have grown considerably in size
and tend to have elongated shapes. Finally, in case 3 where {\it all}
interactions are of the IR type, the matrix particles are fully
included in the clusters formed by the fluid particles; in addition,
the clusters have increased considerably in size compared to the pure
fluid (cf. top panel of figure~\ref{fig:snapshots_A}).

We now perform a similar inspection of simulation snapshots for
systems located in the $(\rho_{\rm f}, \rho_{\rm m})$-plane along path $B$ specified in figure~\ref{fig:pathways}; results are depicted in
figure~\ref{fig:snapshots_B}. Now the {\it total} density, $\rho_{\rm
  t} = \rho_{\rm f} + \rho_{\rm m}$, is kept fixed and we increase
$\rho_{\rm m}$ from 0 to 0.1, i.e., we decrease $\rho_{\rm f}$ from
0.2 to 0.1. Again, along this path $\rho_{\rm m} \leqslant 0.1$, i.e. the
matrix particles are not capable of forming clusters at this
temperature. Similar to the case of systems located along path $A$, we start in the top panel of
figure~\ref{fig:snapshots_B} with a typical equilibrium configuration
of the fluid (i.e., $\rho_{\rm m} = 0$ and hence $\rho_{\rm f} =
0.2$). For the snapshots of the left column we have chosen $\rho_{\rm
  m} = 0.035$ and hence $\rho_{\rm f} = 0.165$. As we immerse the
fluid particles into the matrix formed by HS particles (case 1), we
observe that the clusters formed by the fluid particles populate
preferentially the space left void by the matrix particles, including
in a few cases the essentially inert matrix particles, if imposed by
space requirements. As we now turn on the IR-tail first in the
fluid-matrix and then in the matrix-matrix interactions, the matrix
particles become nucleation centers of the emerging clusters formed by
the fluid particles. In particular in case 3, where the matrix
particles also interact via an IR interaction, the clusters are formed both
by the fluid and the matrix particles, i.e., we observe a microphase
which essentially corresponds to the pure fluid case (see top
panel). As we repeat a similar gedanken-experiment at $\rho_{\rm m} =
0.1$ (and hence $\rho_{\rm f} = 0.1$), we observe the following
microphase formation scenario: in case 1, where the matrix particles
are simple HS, the fluid particles {\it do} form clusters; at
first sight this is  quite astonishing, since -- according to the phase diagram
depicted in figure~4 of~\cite{Sch10} -- at a fluid of density
$\rho_{\rm f} = 0.1$ and at the temperature considered in this
contribution, a pronounced cluster formation is rather
unlikely. However, the rather elevated matrix density essentially reduces
 the available space so that the {\it effective} fluid
density is larger than the nominal fluid density $\rho_{\rm f} = 0.1$,
forcing thereby the fluid particles to form clusters. As we proceed to
case 2, i.e., as we switch on the IR-tail in the fluid-matrix
interaction, microphase formation is strongly suppressed: obviously,
under the effect of the external potential exerted by the matrix
particles, the fluid particles are not capable of forming clusters. Finally,
in case 3 we return to a pronounced cluster-microphase formation where
the matrix particles represent nucleation centers for the fluid
particles.

\subsection{Static structure factors}
\label{subsec:static_structure}

We now examine our data on a more {\it quantitative} level and start
to discuss the results obtained for the static structure factor,
$S_{\rm ff}(k)$, providing information on the correlations between
the fluid particles. $S_{\rm ff}(k)$ has been calculated during the
simulation run according to the following expression
\begin{equation}
S_{\rm ff}(k) =
\frac{1}{N_{\rm f}} \overline{\left< \rho_{\rm f}({\bf k}) \;
\rho_{\rm f}(-{\bf k}) \right>} ~~~ {\rm with} ~~~
\rho_{\rm f}({\bf k}) = \sum_{j=1}^{N_{\rm f}} \exp(\ri {\bf k} \cdot {\bf r}_{j}) .
\end{equation}
The ${\bf r}_j$ are the positions of the fluid particles and the
vectors ${\bf k}$ are compatible with the square geometry of the
simulation box (see, e.g.~\cite{Sch09}). {As a consequence,
  in the low-$k$ regime the number of ${\bf k}$-vectors available is
  smaller, the structure factors are not as smooth as for larger
  $k$-values.} Here $\left< \cdot \cdot \cdot \right>$ denotes the
thermal average over the degrees of freedom of the fluid particles at
a given matrix configuration while $\overline{\phantom{A} \cdots
  \phantom{A}}$ stands for the average over different but equivalent
matrix configurations.

\begin{figure}[htpb]
\begin{tabular}{m{0.1\textwidth}m{0.35\textwidth}m{0.35\textwidth}}
{\bf case 1} &
\includegraphics[width=0.35\textwidth]
          {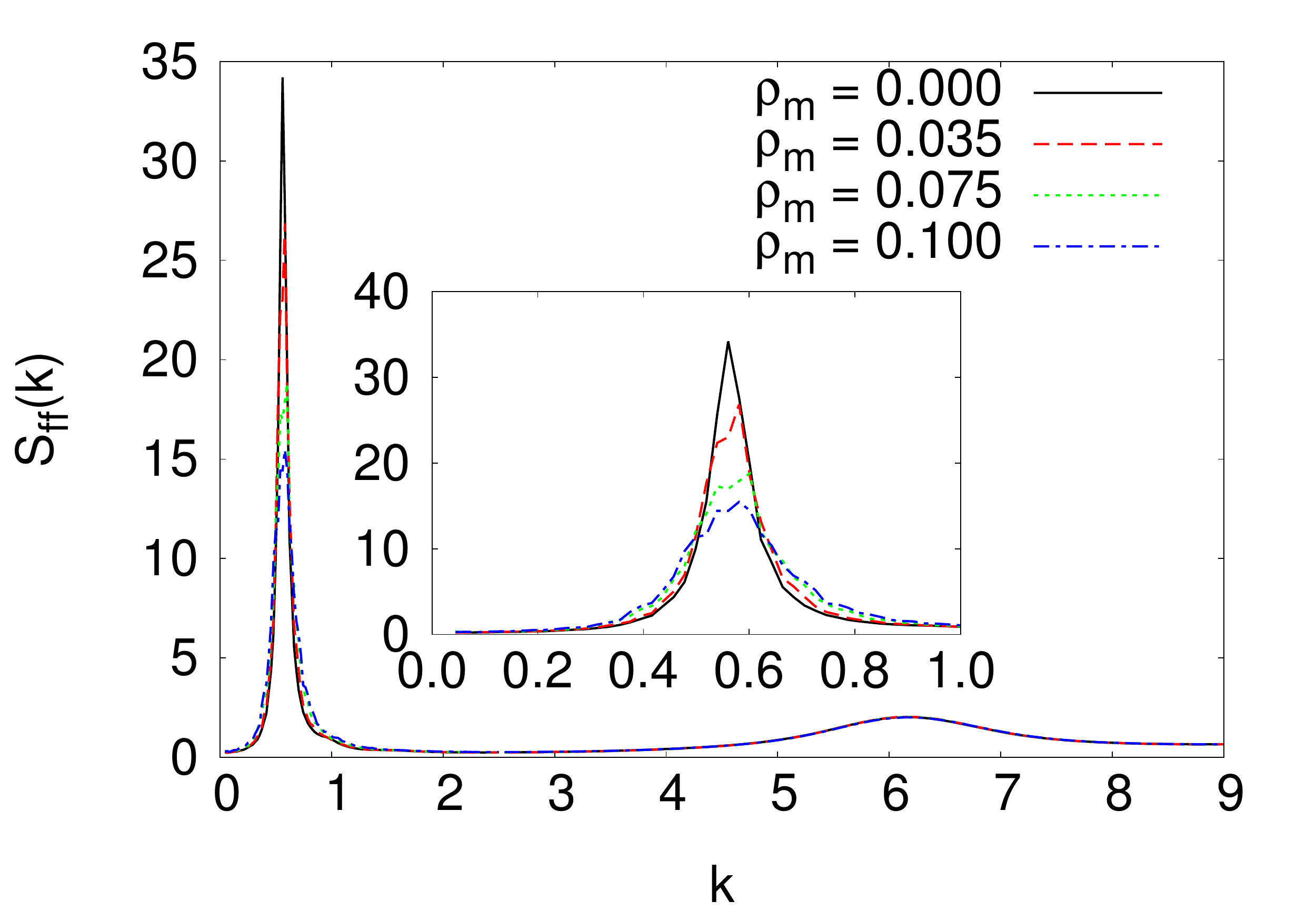} &
\includegraphics[width=0.35\textwidth]
          {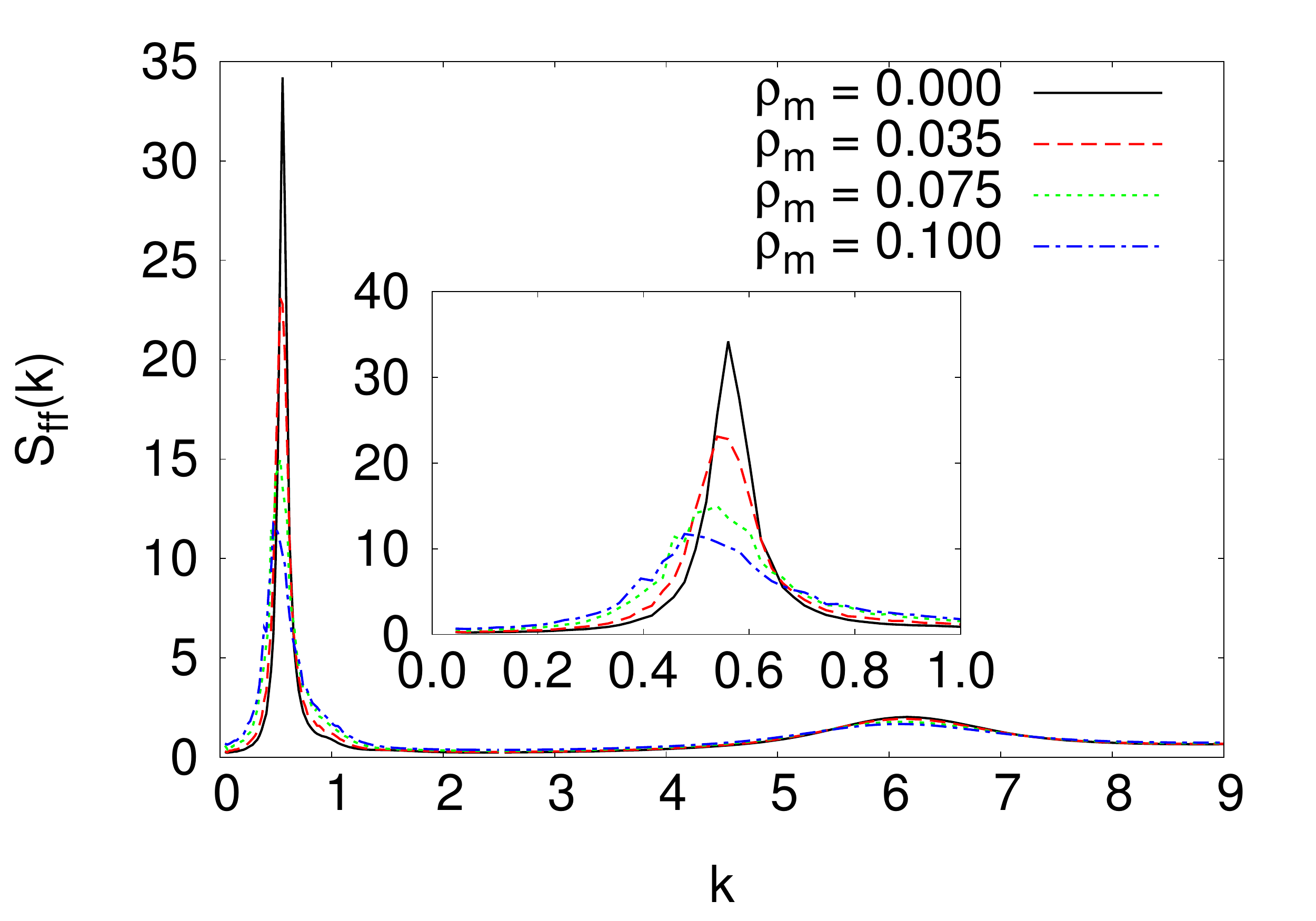} \\
{\bf case 2}&
\includegraphics[width=0.35\textwidth]
          {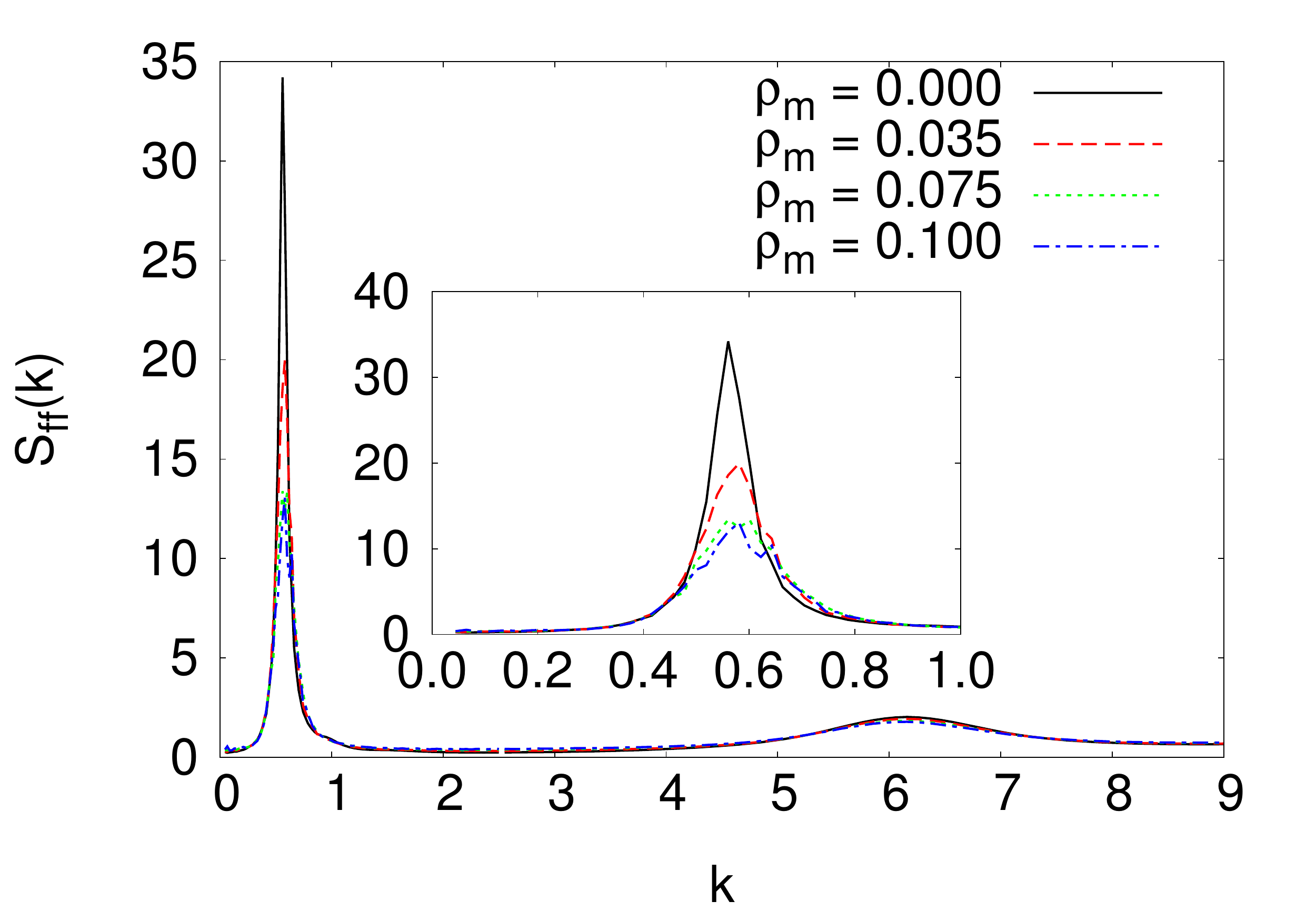} &
\includegraphics[width=0.35\textwidth]
          {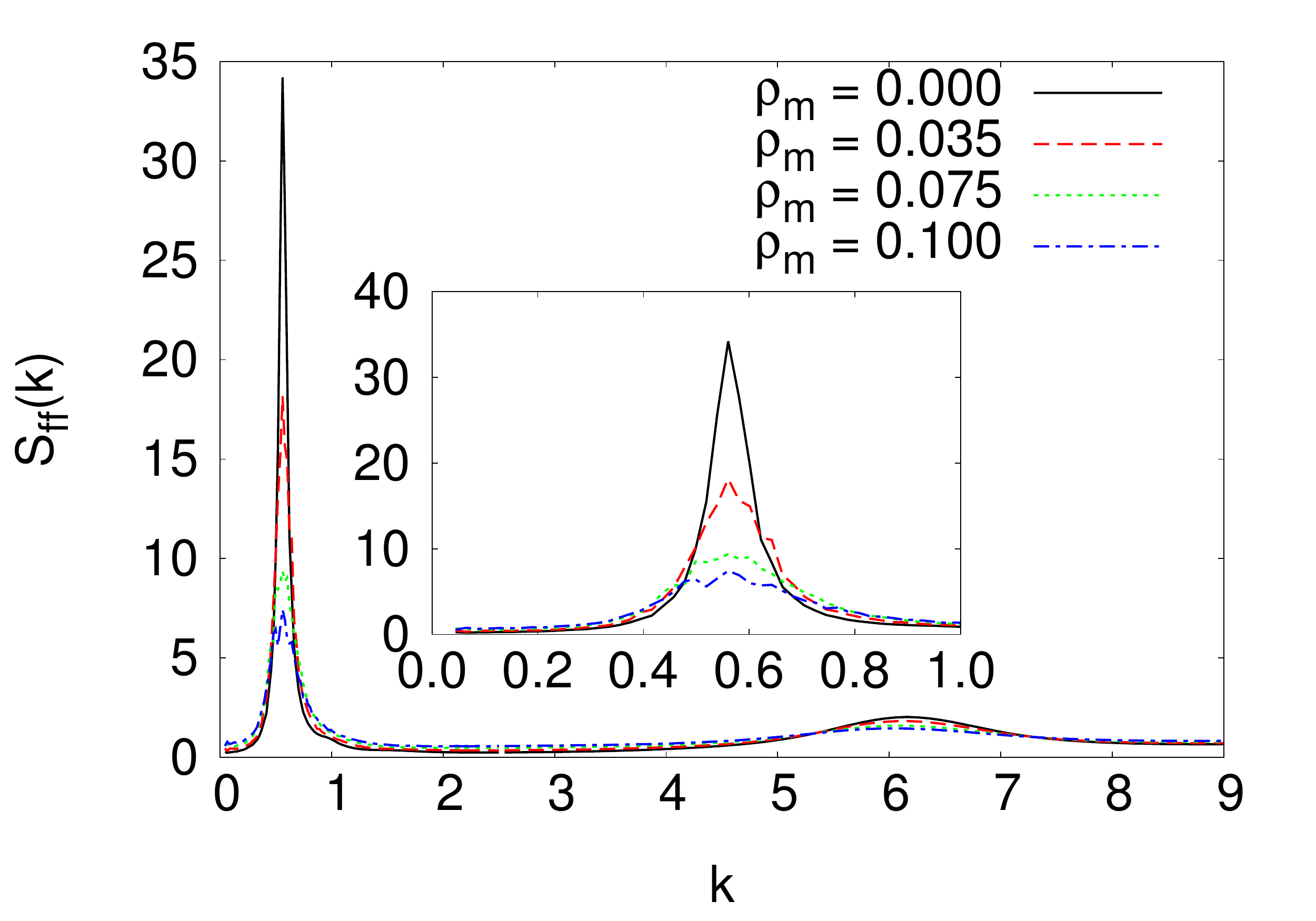} \\
{\bf case 3} &
\includegraphics[width=0.35\textwidth]
          {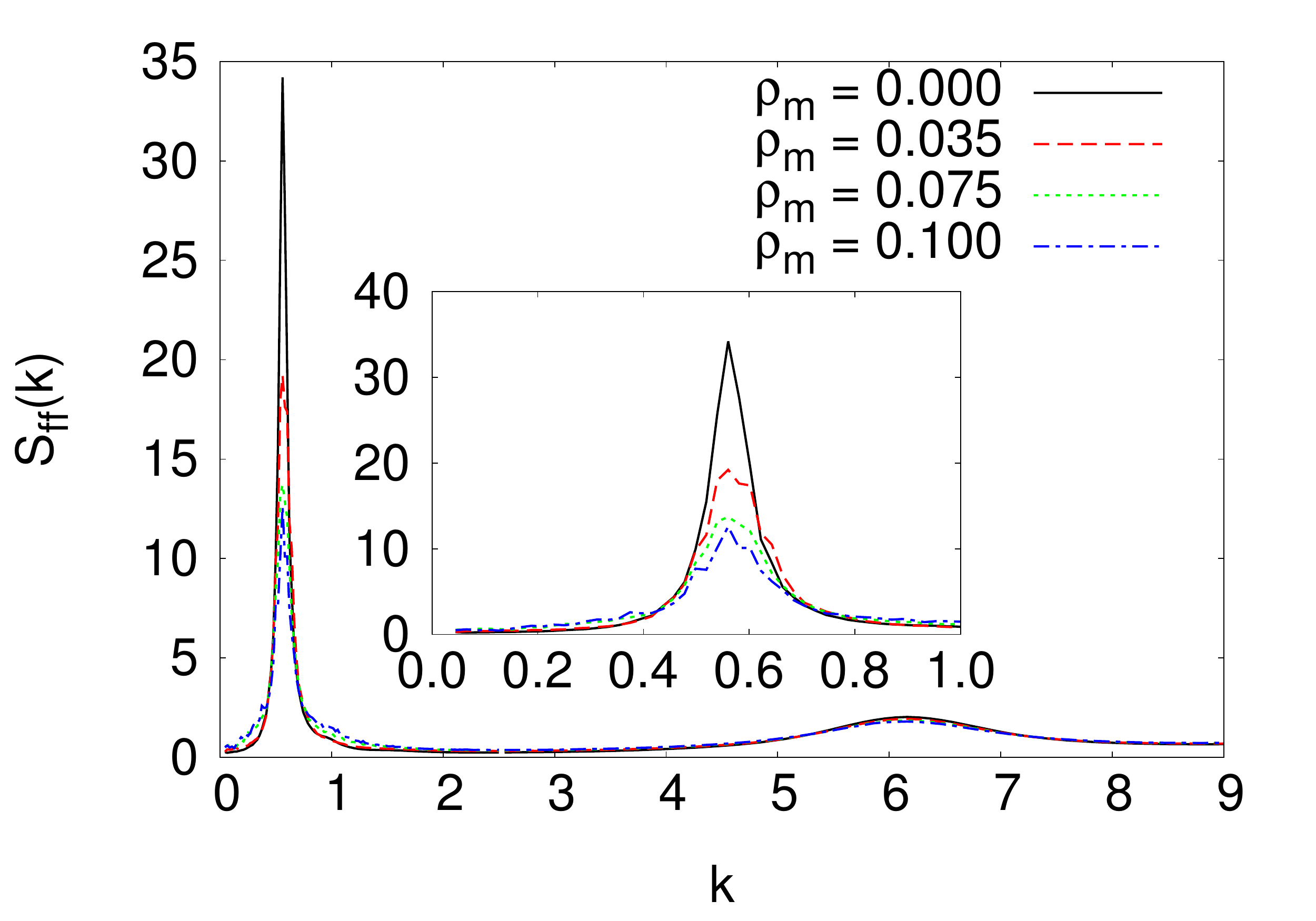} &
\includegraphics[width=0.35\textwidth]
          {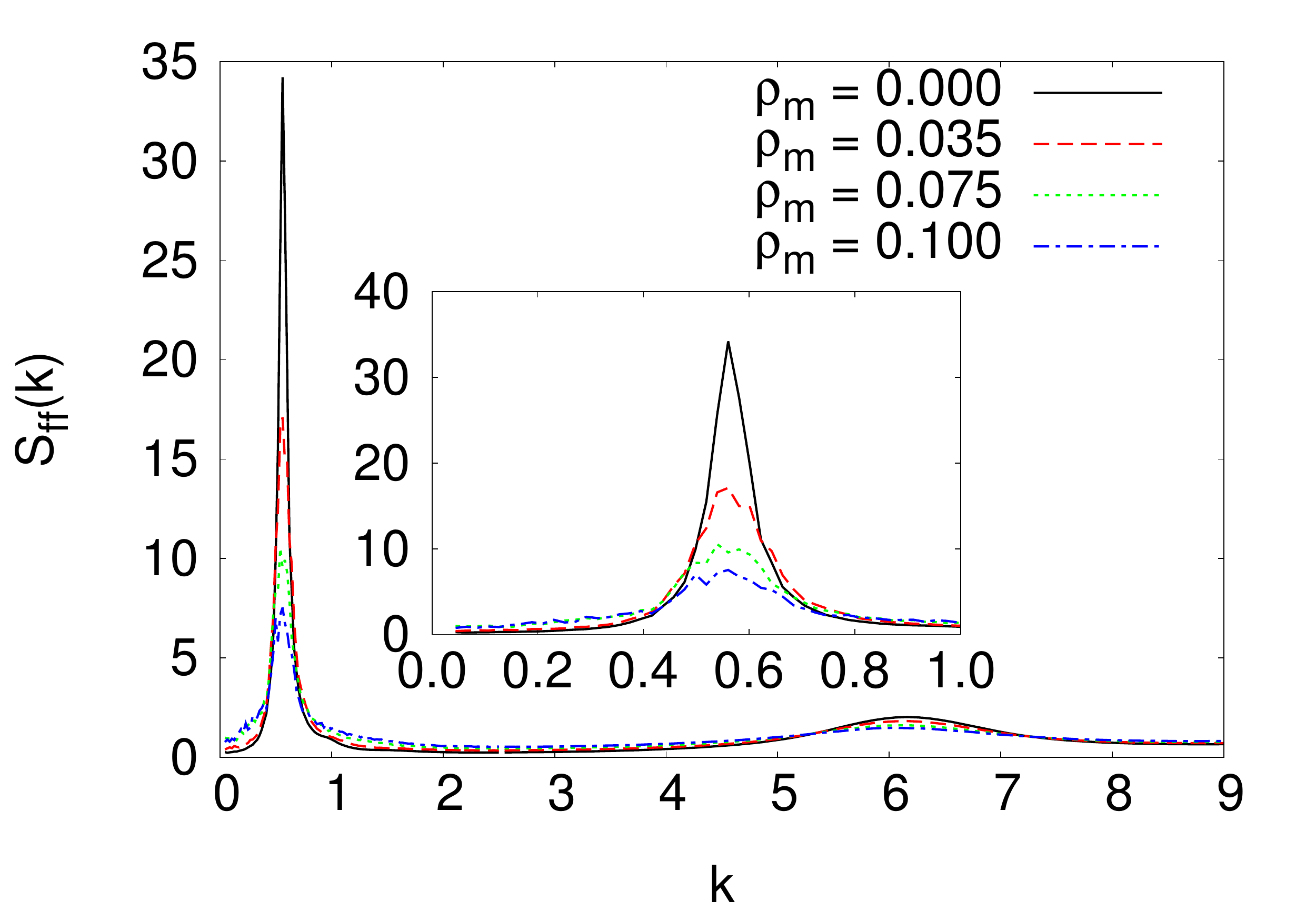} \\
\end{tabular}
\caption{Fluid-fluid static structure factor, $S_{\rm ff}(k)$ as a
  function of $k$ for systems located in the $(\rho_{\rm f}, \rho_{\rm
    m})$-plane along path $A$ (left column) and along path $B$ (right
  column), as specified in figure~\ref{fig:pathways}. The three rows
  correspond to cases 1 to 3, as specified in the text.  Different
  line symbols correspond to different values of $\rho_{\rm m}$ as
  labeled. The insets show enlarged views of the low-$k$ range.}
\label{fig:sff}
\end{figure}
Results for $S_{\rm ff}(k)$ are depicted in figure~\ref{fig:sff}: in
the left (right) column we display the static structure factor for
systems located in the $(\rho_{\rm f}, \rho_{\rm m})$-plane along path $A$ (path $B$), respectively, as defined in figure~\ref{fig:pathways},
considering the three different sets of interactions specified in
section \ref{sec:model_theory}. For reference, the static structure
factor for the {\it pure} fluid ($\rho_{\rm f} = 0.2$ and $\rho_{\rm
  m} = 0$) is shown in all panels.

All structure factors display the same two characteristic features:
they have a rather small peak at $k \sim 6.2$, denoting the {\it
  interparticle} correlations. They show a pronounced first peak at
$k \sim 0.6$, corresponding to the {\it intercluster}
correlations. These two $k$-values provide some rough estimate that
the average interparticle distances are by a factor of ten smaller
than the average intercluster distances. As we increase the matrix
density $\rho_{\rm m}$, we observe throughout a distinct decrease in
the height of the first peak, while the $\rho_{\rm m}$-dependence of
the height of the second peak is rather weak. With respect to the
different interaction scenarios (i.e., cases 1 to 3) we observe
distinct differences in how the height of the first peak decreases as
$\rho_{\rm m}$ is increased: in case 1, where the matrix particles are
essentially inert hard sphere particles, the decay in the peak height is
rather slow with increasing $\rho_{\rm m}$ and thus even at the
highest $\rho_{\rm m}$-value investigated, the intercluster
correlations are still quite pronounced. As we turn on the tail in the
matrix-matrix interactions (case 2) and, finally, in the fluid-matrix
interactions (case 3), the height of the first peak decreases
much faster as $\rho_{\rm m}$ increases, reflecting a
rapid decrease in the correlations between clusters due to the
presence of matrix particles. As we now proceed to systems
located in the $(\rho_{\rm f}, \rho_{\rm m})$-plane along path $B$, we observe
the following scenario. { The tendencies in the decay of
  the first peak with increasing $\rho_{\rm m}$ (and,
  consequently, with decreasing $\rho_{\rm f}$), while we also proceed
  from case 1 to 3, are similar to the ones reported for path $A$. Furthermore we observe that {\it only} for case 1 the positions
  of the first peak shift to lower $k$-values, as $\rho_{\rm m}$
  increases, indicating an increase in the intercluster distance.
  We
  interpret these observations as follows: in case 1 (pure HS matrix),
  the fluid particles, interacting via IR potentials, can form
  well-defined clusters at low $\rho_{\rm m}$-values since the matrix
  particles do not represent an essential spatial restriction
  (cf. corresponding snapshots in figure~\ref{fig:snapshots_B}).
  However, at an increased matrix density,
  less space is left available for the fluid particles to form
  clusters, which are now forced to emerge in the reduced space left
  free by the matrix particles: clusters are now considerably more
  dispersed than for small $\rho_{\rm m}$-values and
  \begin{figure}[!h]
\begin{tabular}{m{0.1\textwidth}m{0.35\textwidth}m{0.35\textwidth}}
{\bf case 1} &
\includegraphics[width=0.35\textwidth]
          {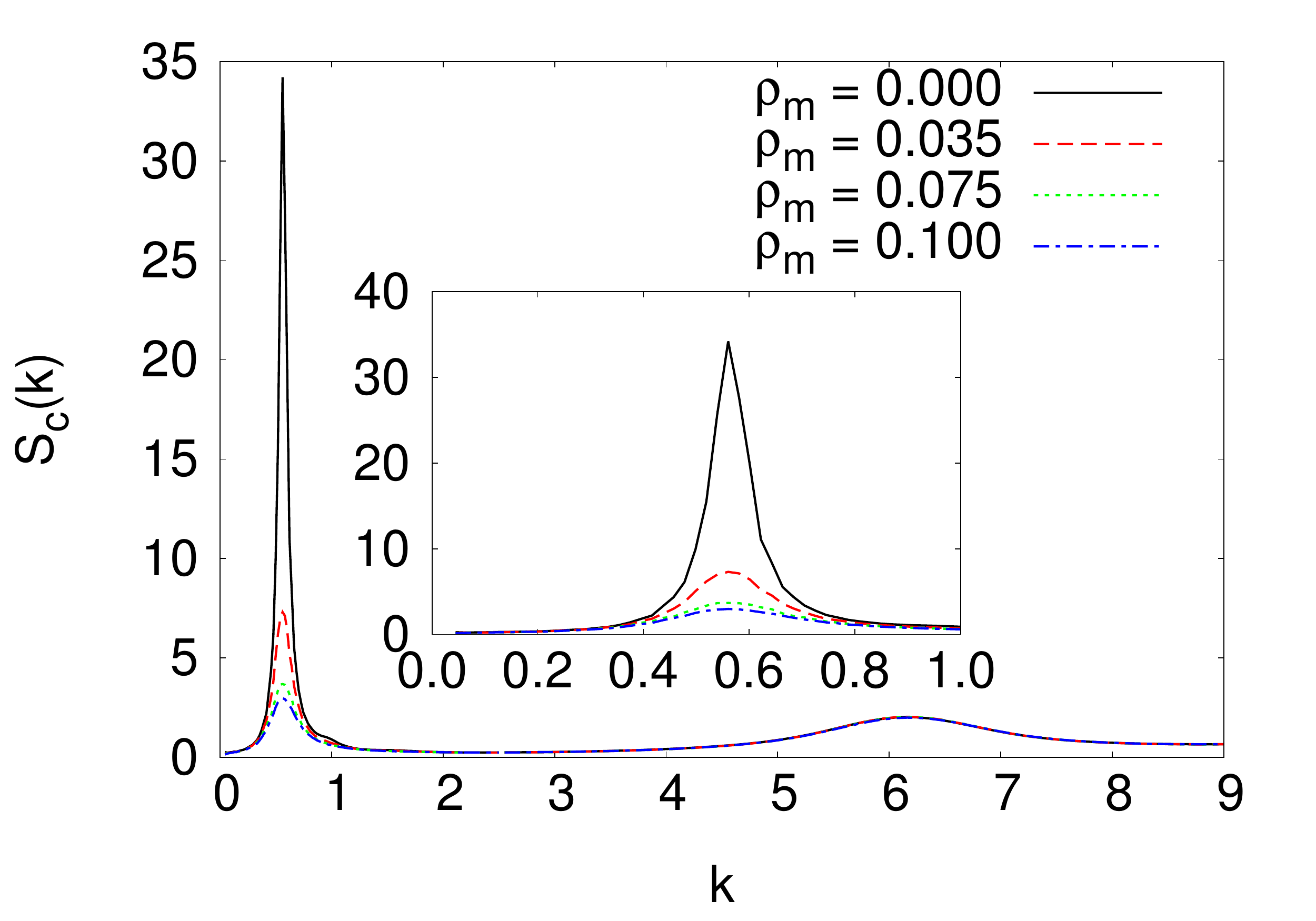} &
\includegraphics[width=0.35\textwidth]
          {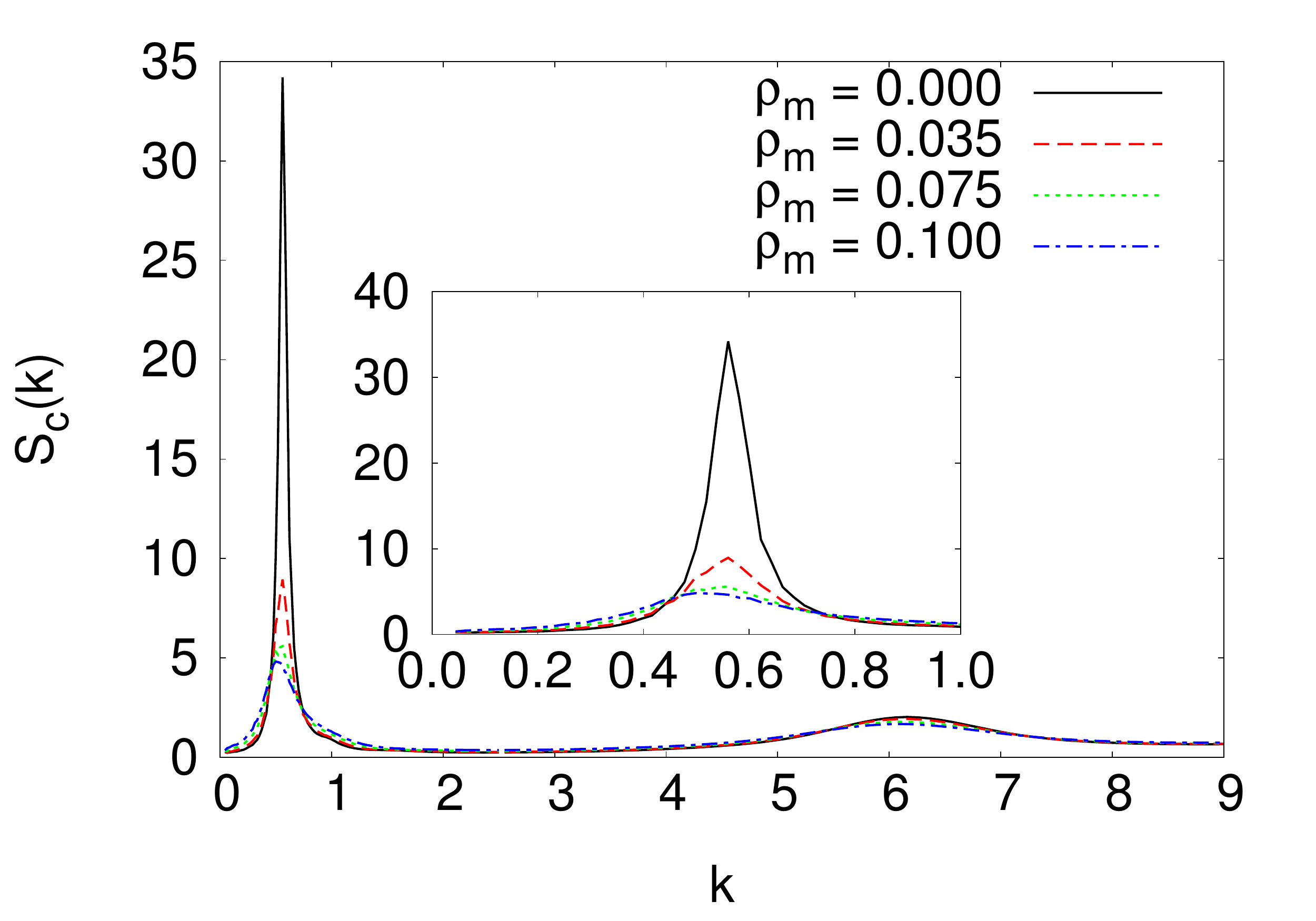} \\
{\bf case 2} &
\includegraphics[width=0.35\textwidth]
          {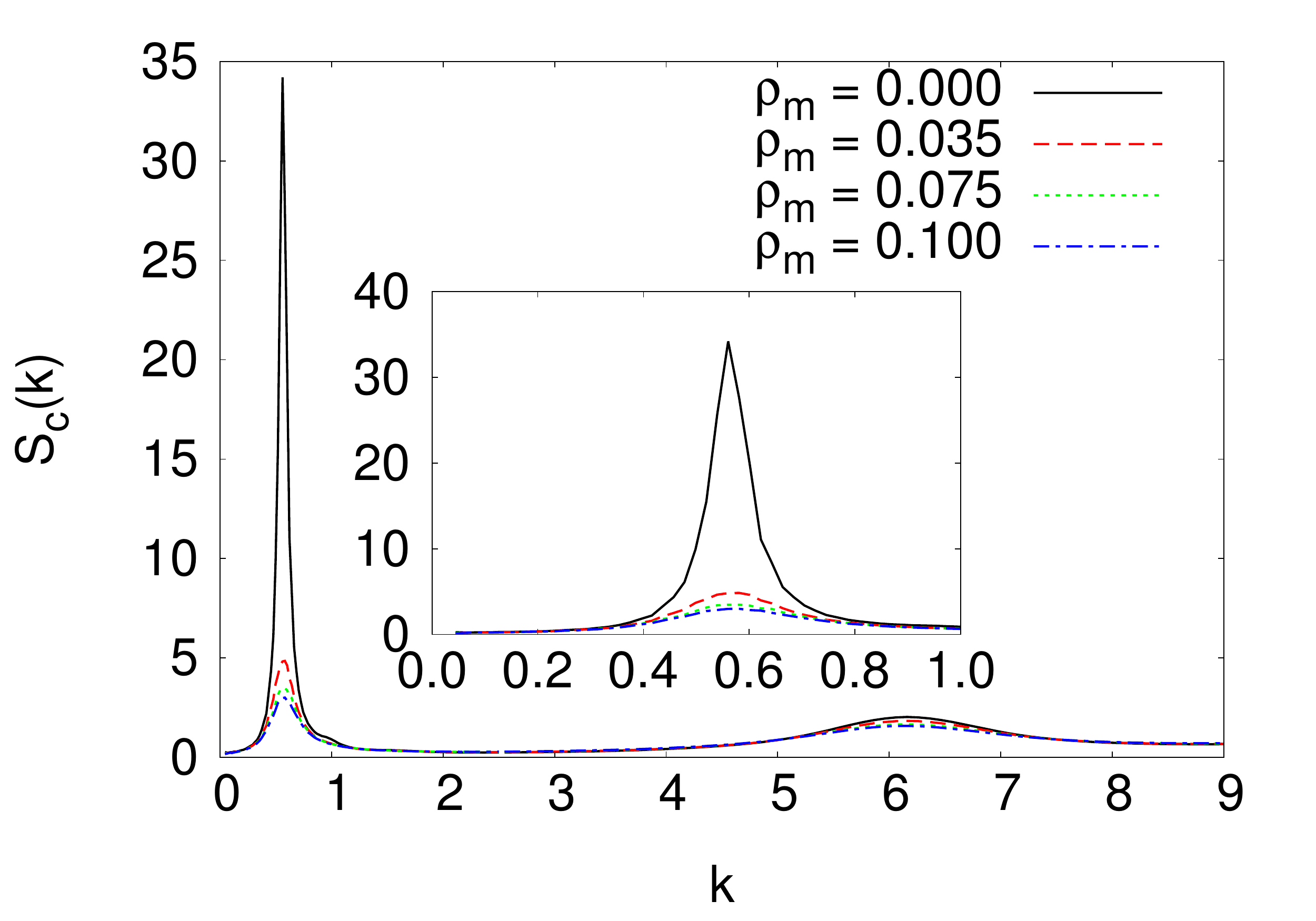} &
\includegraphics[width=0.35\textwidth]
          {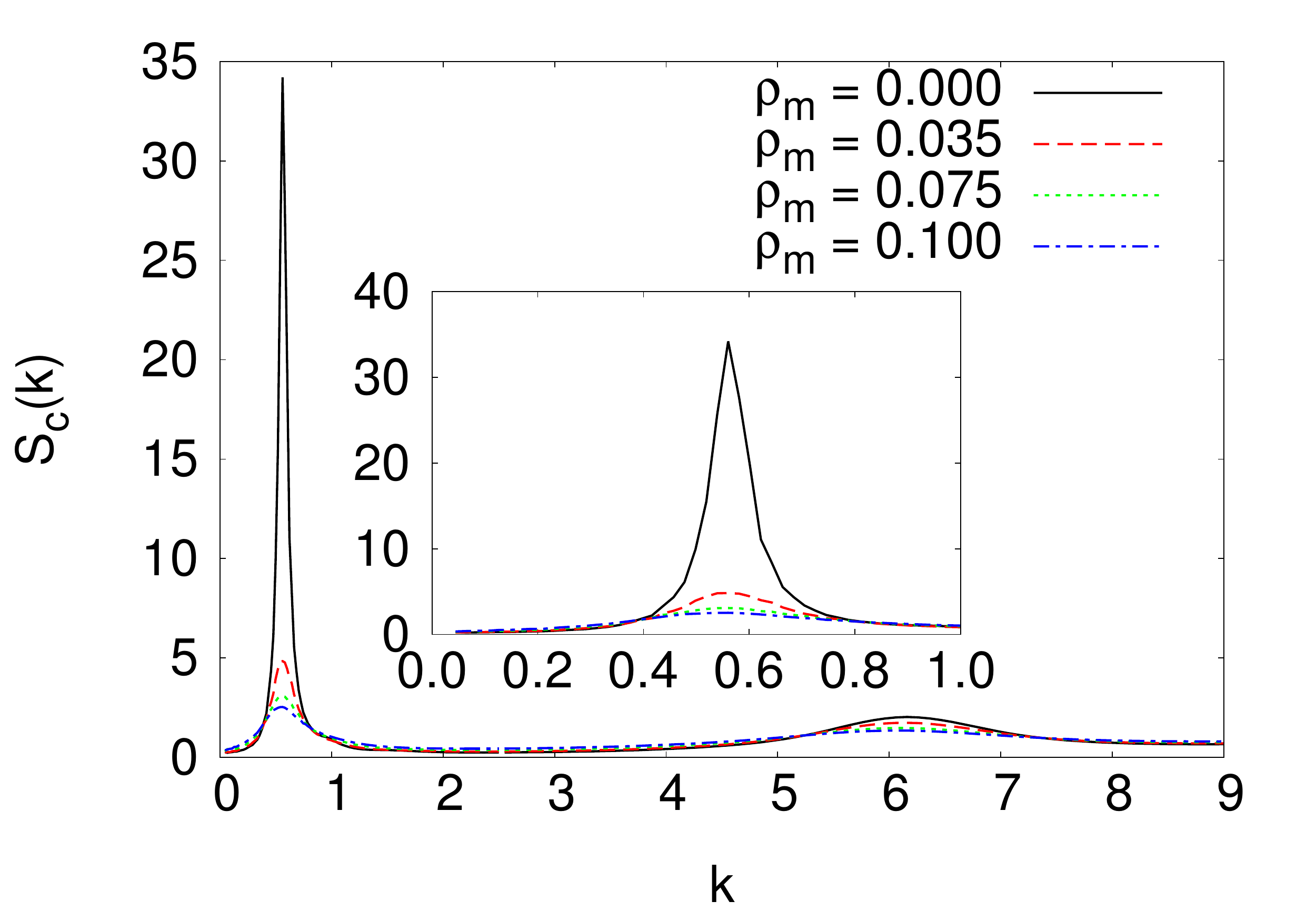} \\
{\bf case 3} &
\includegraphics[width=0.35\textwidth]
          {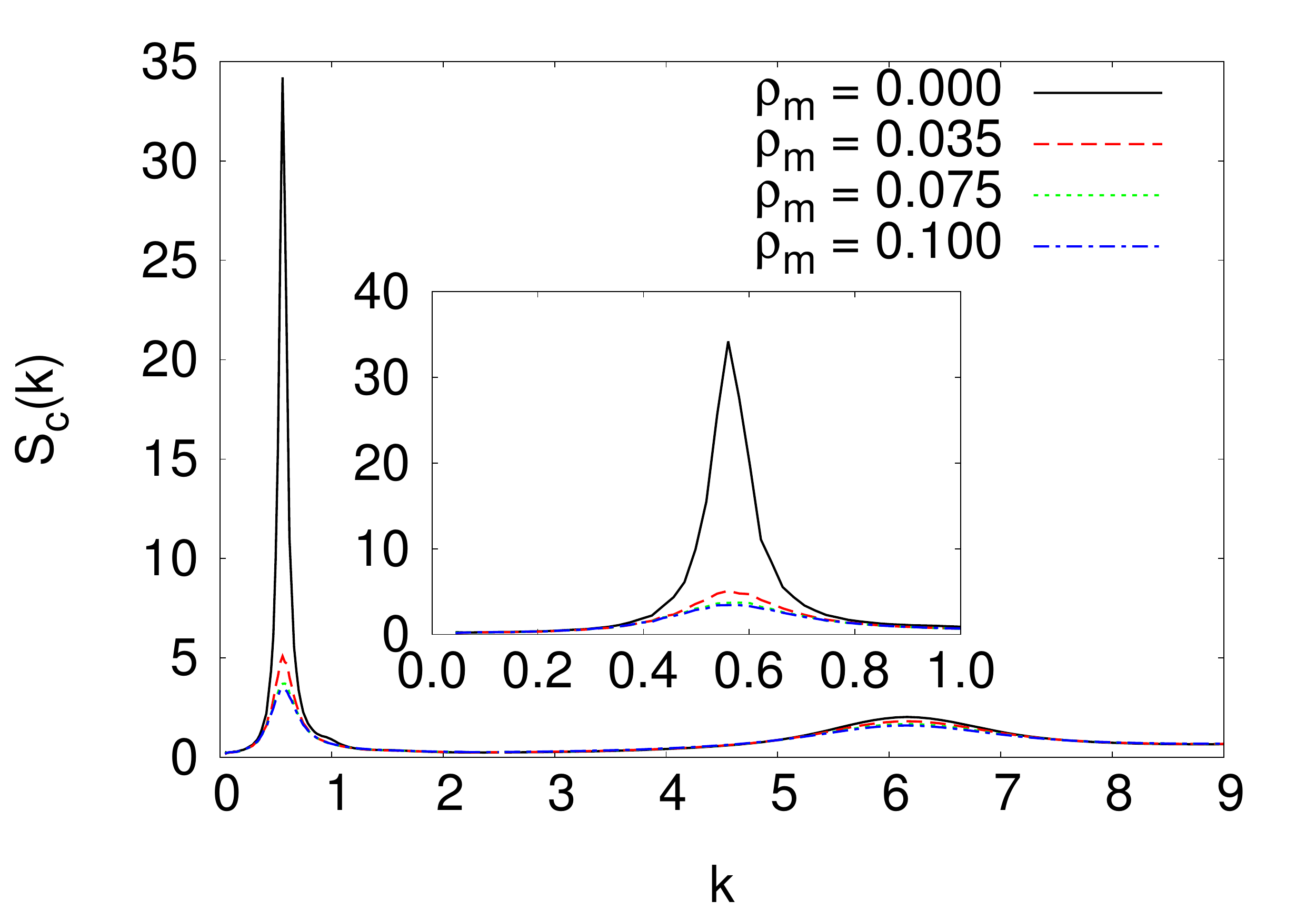} &
\includegraphics[width=0.35\textwidth]
          {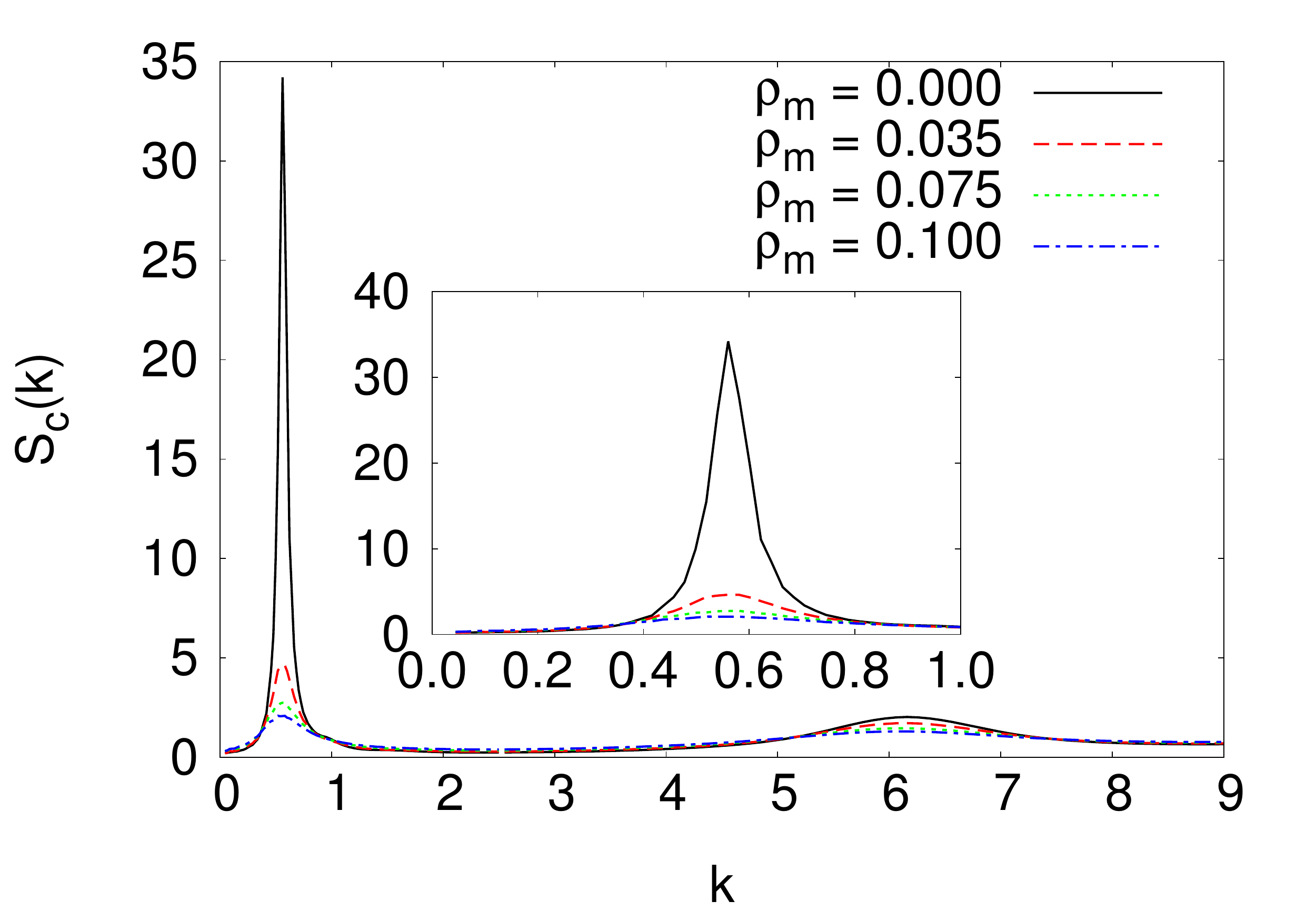} \\
\end{tabular}
\caption{Connected structure factor, $S_{\rm c}(k)$ as a function of
  $k$ for systems located in the $(\rho_{\rm f}, \rho_{\rm m})$-plane
  along path $A$ (left column) and along path $B$ (right column), as
  specified in figure~\ref{fig:pathways}. The three rows correspond to
  cases 1 to 3, as specified in the text.  Different line symbols
  correspond to different values of $\rho_{\rm m}$ as labeled. The
  insets show enlarged views of the low-$k$ range.}
\label{fig:sc}
\end{figure}
are smaller in
  size (again, cf. the corresponding snapshots in figure~\ref{fig:snapshots_A}).
  The situation is different for cases 2 and 3
  as an IR-tail is switched on in the fluid-matrix and/or in the
  matrix-matrix interactions: the attractive components in these
  interaction tails foster the formation of a cluster of the fluid
  particles. This is possible in case 2 despite the presence of the
  matrix particles even at higher $\rho_{\rm m}$-values while in case 3
  the matrix particles are nucleation centers of the clusters formed
  by the fluid particles. Thus, in both cases the intercluster
  distance remains essentially unchanged upon increasing $\rho_{\rm
    m}$, and consequently the position of the first peak does not
  change with $\rho_{\rm m}$ increasing.}

We have also evaluated the connected structure factor, $S_{\rm c}(k)$,
from the particle positions defined via
\begin{equation}
S_{\rm c}(k) =
\frac{1}{N_{\rm f}} \overline{\left< \delta \rho_{\rm f}({\bf k}) \;
\delta \rho_{\rm f}(- {\bf k})\right>}\,,
\end{equation}
where $\delta \rho_{\rm f}({\bf k}) = \rho_{\rm f}({\bf k}) - \left<
\rho_{\rm f}({\bf k}) \right>$. This structure factor provides
information about the correlations between the fluid particles that
are {\it not} mediated via the matrix particles. Results are shown in
figure~\ref{fig:sc}: in the left (right) column we display the static
structure factor for systems located in the $(\rho_{\rm f}, \rho_{\rm
  m})$-plane along path $A$ (path $B$), respectively, as defined in figure~\ref{fig:pathways}, considering three different sets of interactions
specified in section \ref{sec:model_theory}. For reference, the static
structure factor for the pure fluid ($\rho_{\rm f} = 0.2$), to which
$S_{\rm c}(k)$ reduces for $\rho_{\rm m} = 0$, is shown in all panels.

While the variation of the height of the second peak in $S_{\rm c}(k)$
at $k \sim 6.2$ with the densities and the types of interactions shows
the same behaviour as the one reported for $S_{\rm ff}(k)$, we do
observe a distinctively different behaviour on the variation of the
height of the first peak at $k \sim 0.6$ as the system parameters are
changed. Throughout we observe a more pronounced decrease in the
height of this peak as $\rho_{\rm m}$ increases, corresponding to a
gradually increasing suppression of direct correlations between
the fluid particles: at $\rho_{\rm m} = 0.1$, $S_{\rm c}(k \sim 0.6)
\sim 3 (\pm 1)$ while ${S_{\rm ff}(k \sim 0.6) \sim 12 (\pm
  3)}$. In addition, the particular shape of the fluid-matrix
interaction has also an effect on the decrease of the height of the
first peak of the structure factor: the essentially inert HS matrix
particles (case 1) allow for a reasonable amount of direct
correlations between the particles while in case 3, where both fluid
and matrix particles interact via the same type of potential, the
correlations are strongly suppressed.

\subsection{Mean square displacement}
\label{subsec:msd}

\begin{figure}[ht]
\begin{tabular}{m{0.1\textwidth}m{0.35\textwidth}m{0.35\textwidth}}
{\bf case 1} &
\includegraphics[width=0.33\textwidth]
          {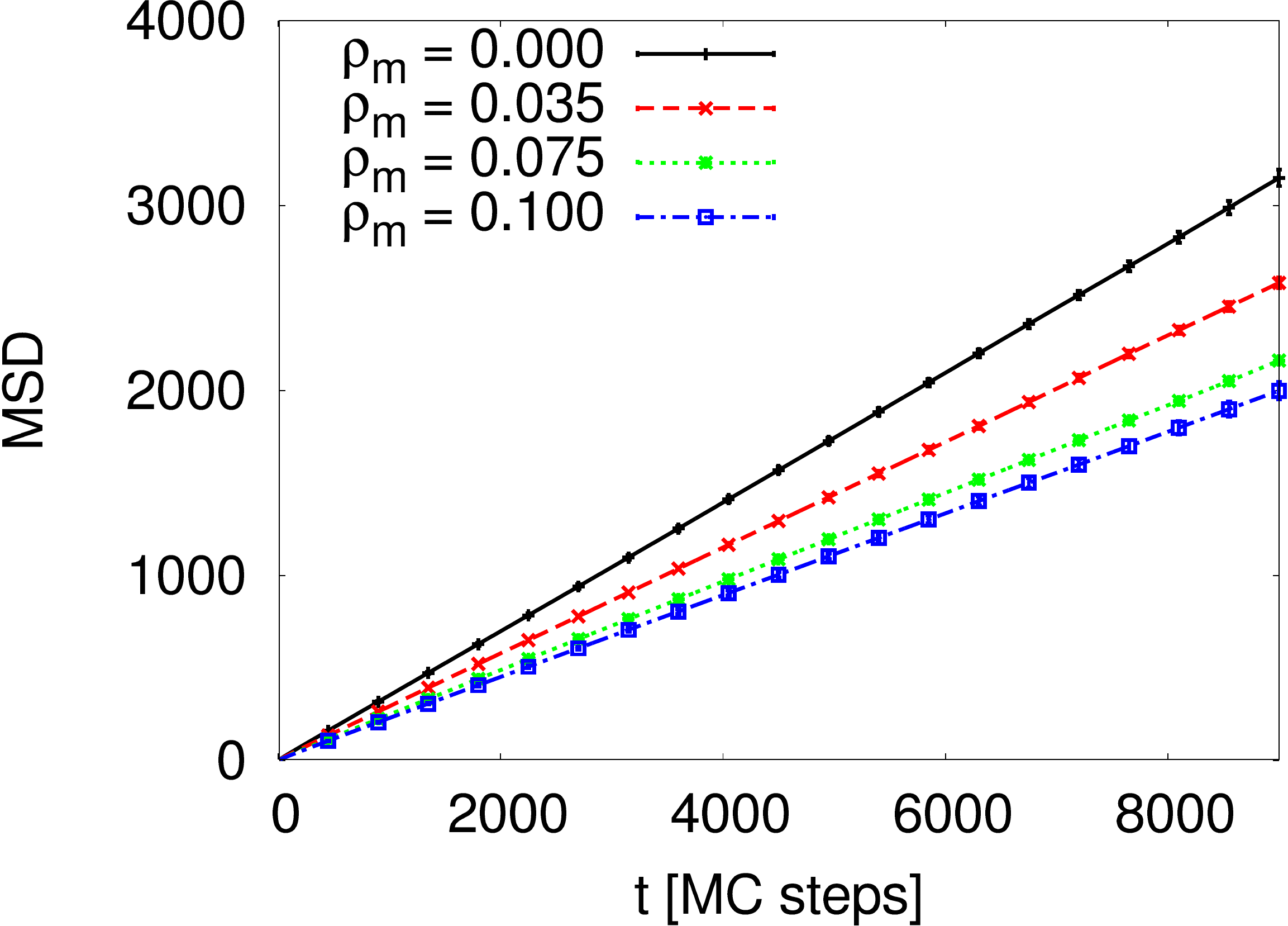} &
\includegraphics[width=0.35\textwidth]
          {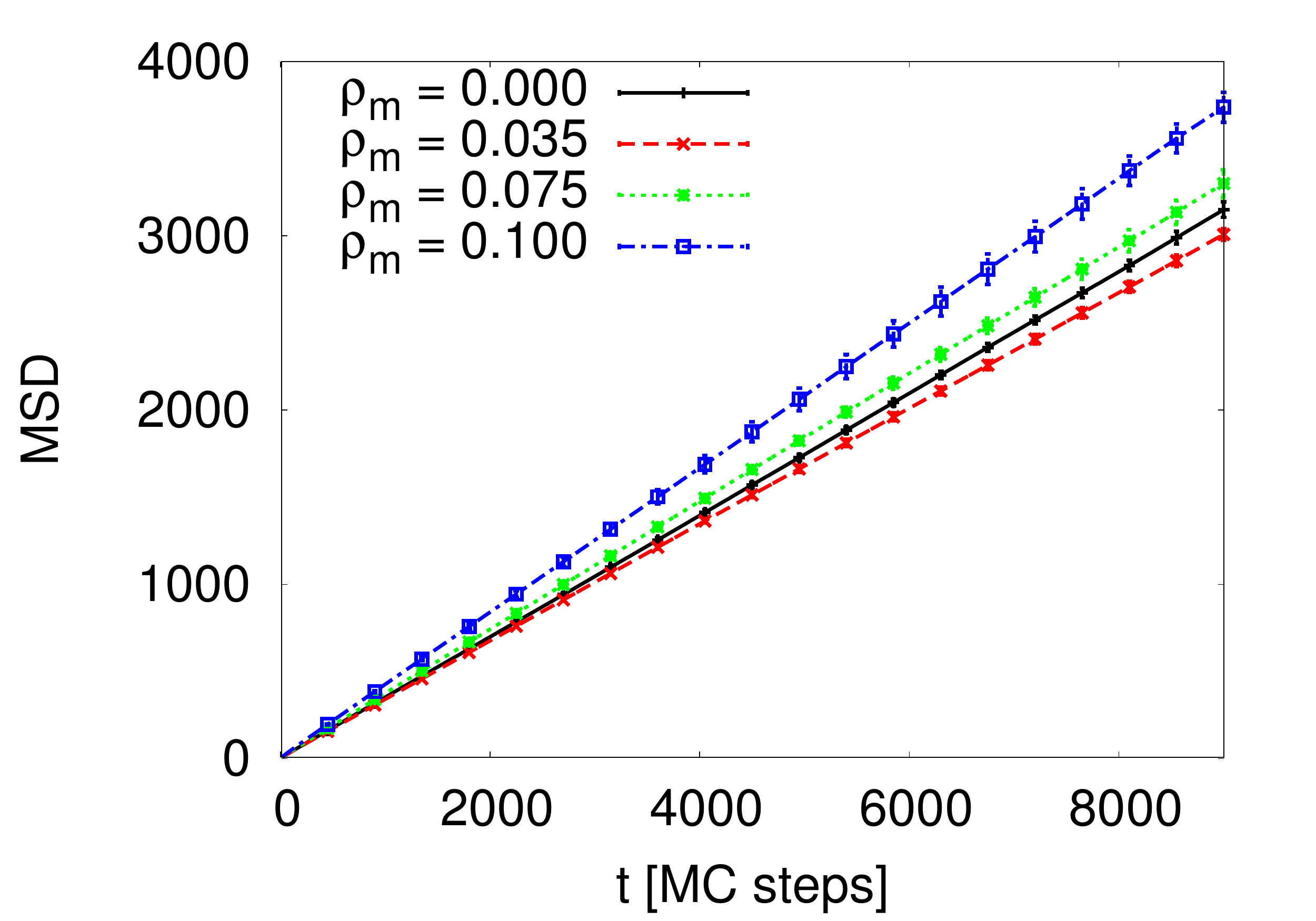} \\
{\bf case 2} &
\includegraphics[width=0.35\textwidth]
          {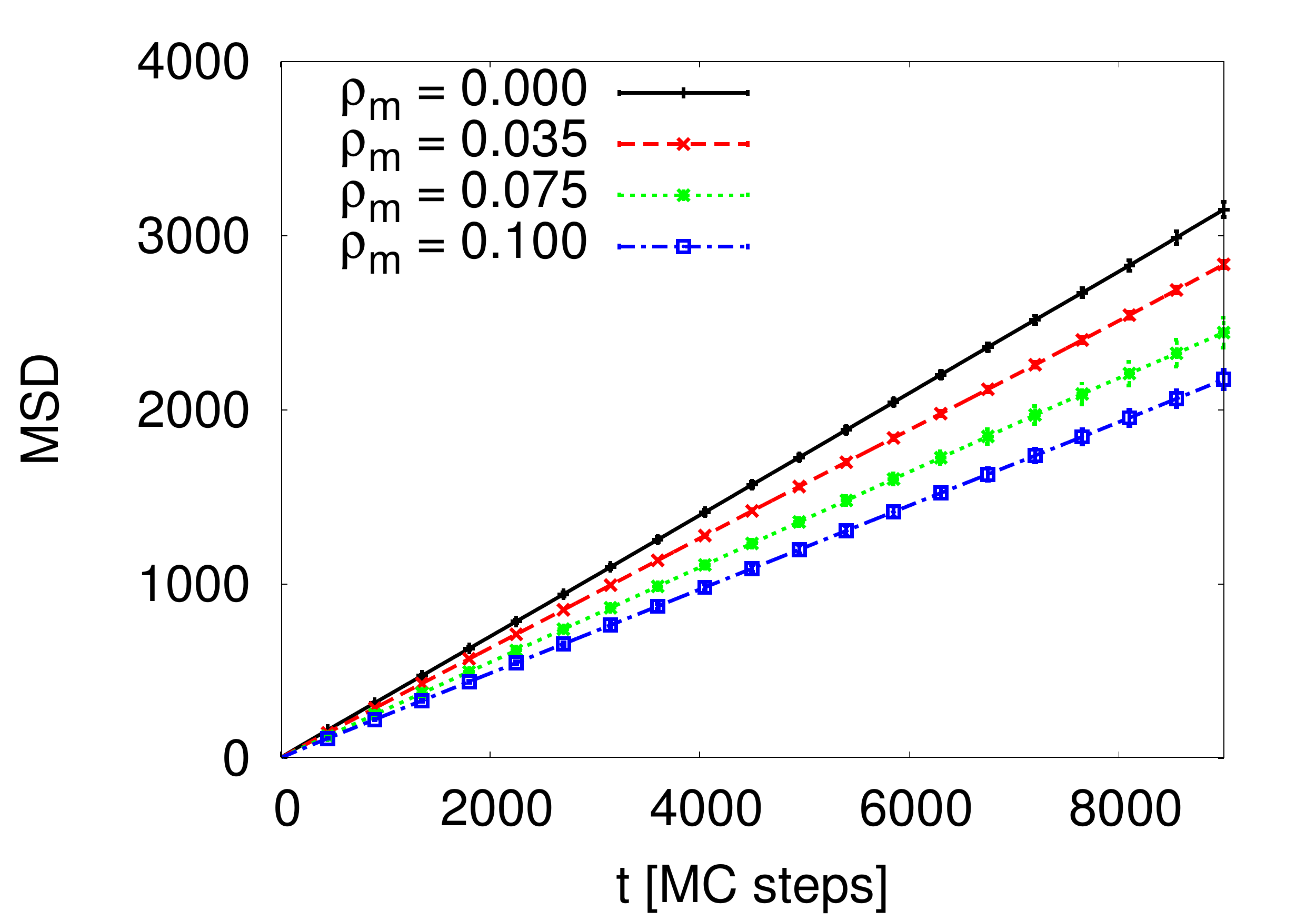} &
\includegraphics[width=0.35\textwidth]
          {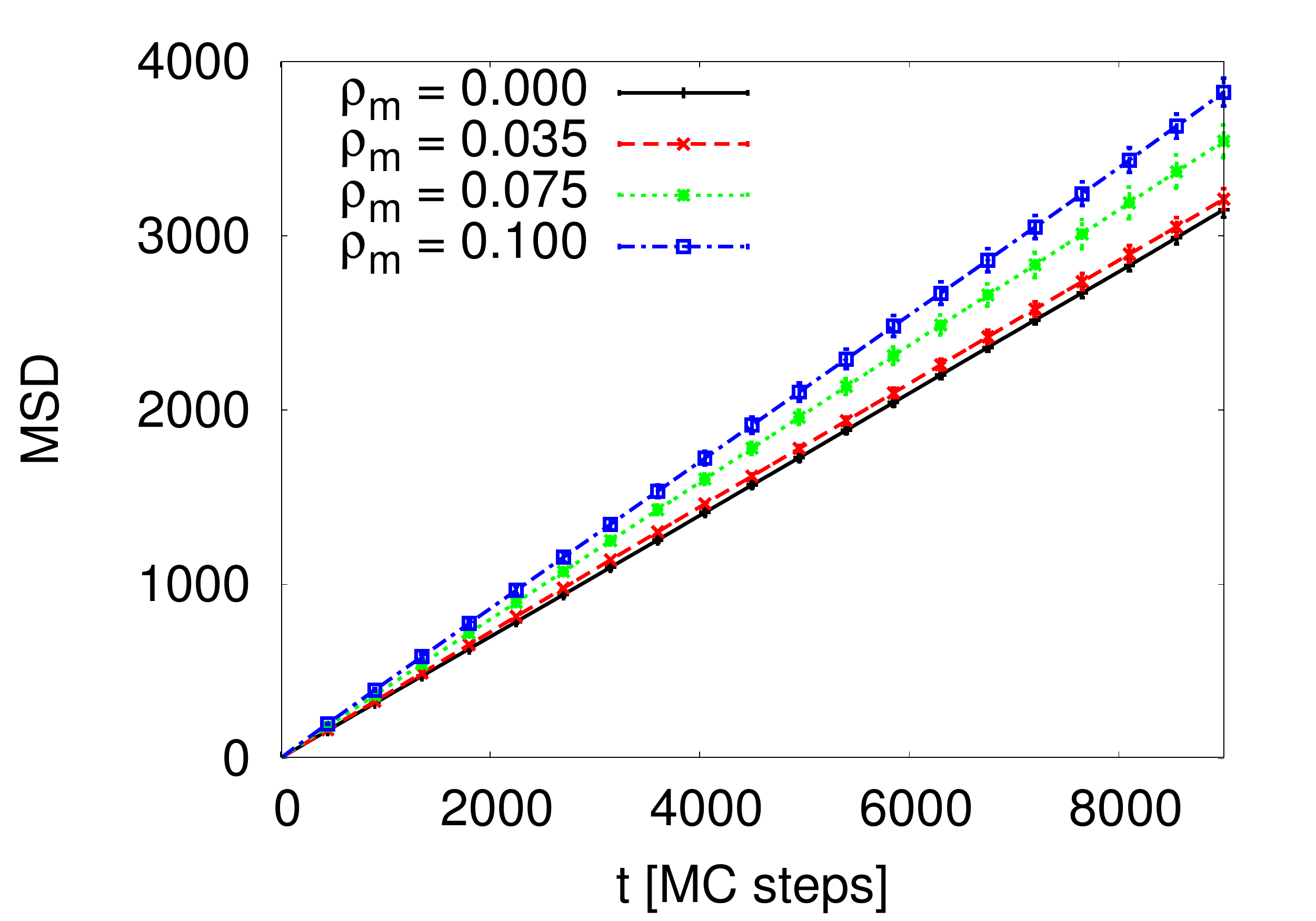} \\
{\bf case 3} &
\includegraphics[width=0.35\textwidth]
          {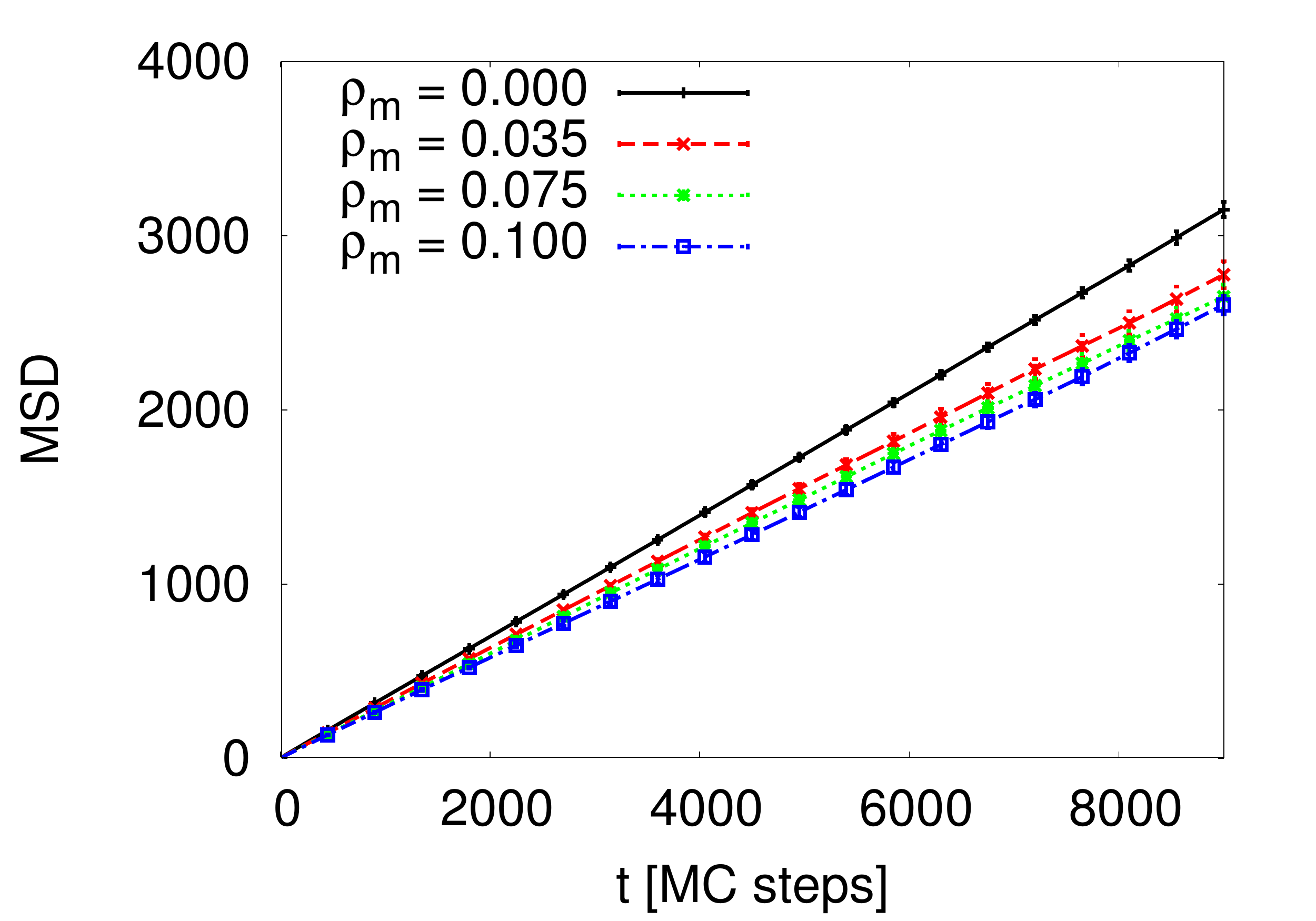} &
\includegraphics[width=0.35\textwidth]
          {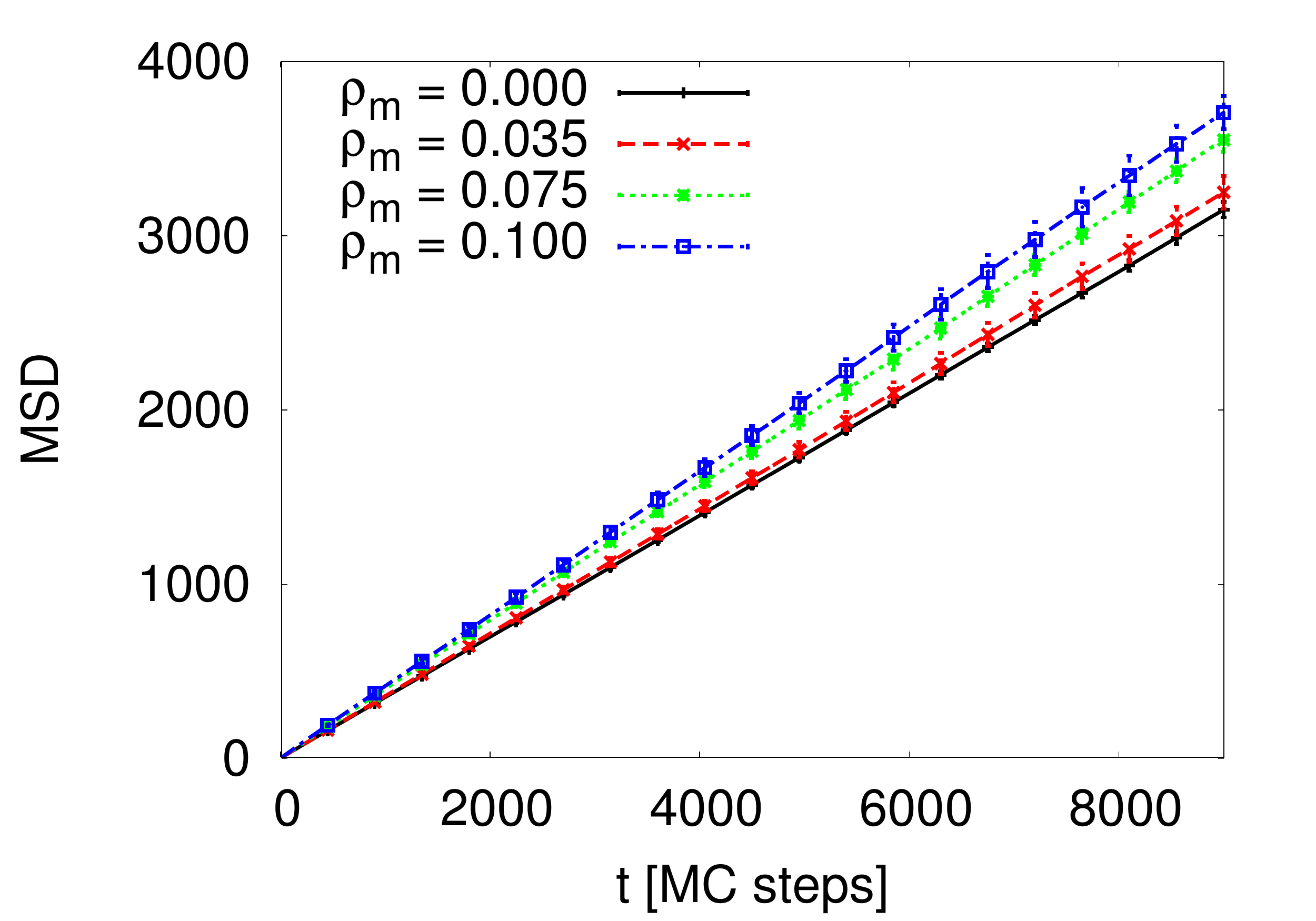} \\
\end{tabular}
\caption{Mean square displacement, $\langle \delta r^2(t) \rangle$, as
  a function of time (measured in terms of Monte Carlo steps, where
  one step corresponds to 100 Monte Carlo sweeps) for systems located
  in the $(\rho_{\rm f}, \rho_{\rm m})$-plane along path $A$ (left
  column) and along path $B$ (right column), as specified in figure~\ref{fig:pathways}. The three rows correspond to cases 1 to 3, as
  specified in the text.  Different line symbols correspond to
  different values of $\rho_{\rm m}$ as labeled.}
\label{fig:msd}
\end{figure}
Finally,  in figure~\ref{fig:msd} we display the mean square
displacement
\begin{equation}
\langle \delta r^2(t) \rangle = \langle | {\bf r}(t) - {\bf r}(0) | \rangle ,
\end{equation}
obtained from the positions of the particles over a respective range
in time, measured in terms of Monte Carlo sweeps where one sweep
corresponds to $N_{\rm f}$ attempted displacement moves (see, e.g.,~\cite{Str11}).  For systems located in the $(\rho_{\rm f}, \rho_{\rm
  m})$-plane along path $A$ (cf. figure~\ref{fig:pathways}), we observe
that an increase in the matrix density $\rho_{\rm m}$ leads in all
cases to a decrease in the diffusion constant $D$ which represents the
slope of the essentially linear mean square displacement curves.
{The results indicate that this decrease in $D$ is more
  rapid with $\rho_{\rm m}$ increasing in case 1 than in the other two
  cases; in particular in case 3 the decrease in $D$ is relatively
  small as $\rho_{\rm}$ increases. We think that these observations
  are the result of two competing effects: on the one hand the fluid
  particles are certainly more trapped within the clusters in cases 2
  and 3, while in case 1 the HS matrix particles form inert
  obstacles. On the other hand, the neatly defined clusters in case 2
  and, in particular, in case 3 (cf. corresponding snapshots in figure~\ref{fig:snapshots_A}) offer large void regions of space where the
  fluid particles can propagate more freely than in the corresponding
  case of a pure HS matrix (cf. corresponding snapshots in figure~\ref{fig:snapshots_A}). For the respective cases, the complex interplay of these two effects
  leads to the above mentioned decrease in
  $D$ as $\rho_{\rm m}$ increases.}  For systems located in the
$(\rho_{\rm f}, \rho_{\rm m})$-plane along path $B$, an increasing matrix
density $\rho_{\rm m}$ leads to an increase in the diffusion
constant. This increase is a bit more pronounced in cases 1 and 2
than in case 3. {We point out that in case 1 a distinct
  non-monotonous behaviour of the diffusion constant at small
  $\rho_{\rm m}$-values ($\rho_{\rm m} \lesssim 0.035 $) is
  observed as the matrix density increases. A similar
  behaviour is encountered in cases 2 and 3 only for matrix densities
  up to $\rho_{\rm m} \simeq 0.01$.}

\section{Conclusions}
\label{sec:conclusions}

In this contribution we have investigated the cluster microphase
formation of a two-dimensional fluid annealed in the presence of a
quenched matrix configuration of fluid particles. In all cases
investigated the fluid particles interact via a hard sphere potential
with an adjacent tail of competing interactions. In an effort to study
the cluster formation of the fluid particles under the effect of
the matrix particles we have varied both the fluid and the matrix
densities, $\rho_{\rm f}$ and ${\rho_{\rm m}}$, as well as
the fluid-matrix and the matrix-matrix interactions, ranging from
simple hard spheres to hard spheres with competing tail
interactions. Our investigations are based on extensive Monte Carlo
simulations. The observables, such as static and dynamic correlation
functions, are obtained by a double averaging procedure: one average
is taken over the degrees of freedom of the mobile fluid particles for
a particular matrix realization, the other one is taken over
different but equivalent matrix configurations.

We have discussed the observed phenomena on a {\it qualitative} level
by visual inspection of the selected representative snapshots and on a
{\it quantitative} level by evaluating the static structure [i.e., the
  structure factors $S_{\rm ff}(k)$ and $S_{\rm c}(k)$] as well as the
mean square displacement, $\langle \delta r^2(t) \rangle$. These
results provide clear evidence that the matrix does have a distinct
effect on the microphase formation, which can be summarized as
follows: small matrix densities leave the cluster formation
essentially unaffected while larger $\rho_{\rm m}$-values do effect
the microphase formation. Here the explicit shape of the fluid-matrix
and of the matrix-matrix interactions become relevant: simple hard
sphere matrix particles can effect the cluster formation by the
fact that they are capable of reducing the space available to the fluid
particles, increasing thereby the effective matrix density; on the
other hand, if the interactions between the fluid and the matrix
particles is of the same type, we have found that the matrix particles
act as nucleation centers for the emerging cluster microphase of the
fluid particles.

\section*{Acknowledgements}

Financial support by the Austrian Research Fund (FWF) under
Proj. Nos. P19890--N16 and W004 is gratefully acknowledged.  The
authors are indebted to Daniele Coslovich (Montpellier) and Jan
Kurzidim (Wien) for helpful discussions.

One of us (GK) is in particular grateful to Yura Kalyuzhnyi to whom
this contribution is dedicated: our close cooperation has been
extremely fruitful, both from the scientific and from the
personal point of view.

\newpage
\ukrainianpart

\title{Двовимірні системи із конкуруючими взаємодіями: формування мікрофази під впливом невпорядкованого пористого середовища}

\author{Д.Ф. Шванцер, Г. Каль}

\address{Інститут теоретичної фізики та центр чисельного матеріалознавства, Технічний університет Відня, Відень, Австрія}

\makeukrtitle

\begin{abstract}
Нами досліджено вплив невпорядкованого пористого середовища на утворення кластерної мікрофази у двовимірній системі, у якій мають місце конкуруючі взаємодії між частинками. З цією метою ми здійснили грунтовне моделювання методом Монте-Карло, систематично змінюючи густину плину та середовища, а також взаємодію між частинками середовища і взаємодію між частинками середовища та плину. Наші результати доводять, що середовище істотно впливає на формування мікрофази частинок плину: якщо частинки взаємодіють як між собою так і з частинками плину через звичайний потенціал твердих сфер, то вони, по суті, зменшують доступний об'єм, у якому частинки плину формують кластерну мікрофазу. З іншого боку, якщо розглядати далекосяжну частину взаємодій ``середовище -- середовище'' і ``середовище -- плин'', то частинки середовища стають центрами нуклеації для кластерів, утворених частинками плину.

\keywords м'яка речовина, пористі середовища, утворення мікрофази, статична структура, динамічні властивості
\end{abstract}

\end{document}